\documentclass[%
 reprint,
 amsmath,amssymb,
 aps,
]{revtex4-2}
\usepackage[utf8]{inputenc} 
\DeclareUnicodeCharacter{0229}{\c{e}}
\usepackage{graphicx}
\usepackage{dcolumn}
\usepackage{bm}
\usepackage{here}
\usepackage{amsmath}
\usepackage{amssymb}
\usepackage{amsfonts}
\usepackage{color}
\usepackage{ascmac}
\usepackage{nameref}
\usepackage{physics}
\usepackage{xr}
\newcommand{\argmax}{\mathop{\rm arg~max}\limits}

\usepackage{multibib}
\newcites{SI}{Supplementary References}

\let\origaddcontentsline\addcontentsline
\newif\ifDisableTOC
\DisableTOCfalse
\renewcommand{\addcontentsline}[3]{%
  \ifDisableTOC\else
    \origaddcontentsline{#1}{#2}{#3}%
  \fi
}
\DisableTOCtrue

\begin{document}


\preprint{APS/123-QED}
\title{Koopman Mode Decomposition of Thermodynamic Dissipation \\ in Nonlinear Langevin Dynamics}

\author{Daiki Sekizawa}
\email{sekizawa-daiki963@g.ecc.u-tokyo.ac.jp}
\affiliation{
Department of General Systems Studies, The University of Tokyo, 3-8-1 Komaba, Meguro-ku, Tokyo 153-8902, Japan}

\author{Sosuke Ito}
\email{s-sosuke.ito@g.ecc.u-tokyo.ac.jp}
\affiliation{Department of Physics, The University of Tokyo, 7-3-1 Hongo, Bunkyo-ku, Tokyo 113-0033, Japan}
\affiliation{Universal Biology Institute, The University of Tokyo, 7-3-1 Hongo, Bunkyo-ku, Tokyo 113-0033, Japan}
\author{Masafumi Oizumi}
\email{c-oizumi@g.ecc.u-tokyo.ac.jp}
\affiliation{Department of General Systems Studies, The University of Tokyo, 3-8-1 Komaba, Meguro-ku, Tokyo 153-8902, Japan}
\date{\today}

\begin{abstract}
Nonlinear oscillations are commonly observed in complex systems far from equilibrium, such as living organisms. These oscillations are essential for sustaining vital processes, like neuronal firing, circadian rhythms, and heartbeats. In such systems, thermodynamic dissipation is necessary to maintain oscillations against noise. However, due to their nonlinear dynamics, it has been challenging to determine how the characteristics of oscillations, such as frequency, amplitude, and coherent patterns across elements, influence dissipation. To resolve this issue, we employ Koopman mode decomposition, which recasts nonlinear dynamics as a linear evolution in a function space. This linearization allows the dynamics to be decomposed into temporal oscillatory modes coherent across elements, with the Koopman eigenvalues determining their frequencies. Using this method, we decompose thermodynamic dissipation caused by nonconservative forces into contributions from oscillatory modes in overdamped nonlinear Langevin dynamics. We show that the dissipation from each mode is proportional to its frequency squared and its intensity, providing an interpretable, mode-by-mode picture. In the noisy FitzHugh--Nagumo model, we demonstrate the effectiveness of this framework in quantifying the impact of oscillatory modes on dissipation during nonlinear phenomena like {\color{black} coherent resonance} and bifurcation. For instance, our analysis of {\color{black} coherent resonance} reveals that the greatest dissipation at the optimal noise intensity is supported by a broad spectrum of frequencies, whereas at non-optimal noise levels, dissipation is dominated by specific frequency modes. Our work offers a general approach to connecting oscillations to dissipation in noisy environments and improves our understanding of diverse oscillation phenomena from a nonequilibrium thermodynamic perspective. 
\vspace{6pt} 
\end{abstract}

\maketitle

\section{Introduction}
Oscillations are a pervasive phenomenon in nature. Examples range from the rhythmic beating of the heart~\cite{van1928lxxii, noble1962modification} and circadian clocks that regulate sleep-wake cycles~\cite{konopka1971clock, pittendrigh1960circadian}, to neuronal firing patterns~\cite{Buzsaki2006-se} and periodic chemical waves in the Belousov–Zhabotinsky reaction~\cite{Belousov1958-yu, zhabotinsky1964periodic}. These oscillations require physical processes to be far from thermodynamic equilibrium in order to persist~\cite{sekizawa2024decomposing, barato2017coherence, Del-Junco2020-px, Oberreiter2022-nx,ohga2023thermodynamic, shiraishi2023entropy, dechant2023thermodynamic, aguilera2025inferring, aslyamov2025excess,
Barato2016-lh, Cao2015-ho, Fei2018-ae, Lee2018-kb, nardini2017entropy, santolin2025dissipation, nagayama2025duality, Uhl2019-ch, kolchinsky2024thermodynamic,zheng2024topological,  xu2025thermodynamic}. 
In the steady state, the extent to which a system deviates from thermodynamic equilibrium is quantified by the entropy production rate~\cite{seifert2025stochastic}. The entropy production rate is a nonnegative quantity that measures total irreversibility. It also illustrates the extent to which nonconservative forces violate the detailed balance condition, thereby causing oscillatory behavior in the steady state.

This paper addresses the key question of how the characteristics of oscillations, such as frequency, amplitude, and coherent patterns across elements, determine the entropy production rate.  In a previous paper~\cite{sekizawa2024decomposing}, we attempted to answer the question by considering the eigenmode decomposition of the housekeeping entropy production rate~\cite{Maes2014-pk,Nakazato2021-do, dechant2022geometric, Dechant2022-gt, Ito2024-yt}, which is the amount of dissipation caused only by nonconservative forces. However, the analysis was limited to linear forces and could not account for various oscillatory behaviors~\cite{ strogatz2024nonlinear, gammaitoni1998stochastic, izhikevich2007dynamical,murray2007mathematical}, such as limit cycles, bifurcations, in which oscillations suddenly appear or change character, and {\color{black} coherent resonance}, in which noise paradoxically stabilizes rhythmic activity. Although several studies have examined the relationship between nonlinear phenomena and thermodynamic dissipation ~\cite{
qian2000pumped,vellela2009stochastic, ge2009thermodynamic, falasco2018information,lucarini2019stochastic,  yoshimura2021thermodynamic, 
yan2023thermodynamic, 
remlein2024nonequilibrium, santolin2025dissipation,falasco2025macroscopic, nagayama2025geometric,nagayama2025duality}, the inherent complexity of general nonlinear systems makes it difficult to interpret how their dissipation arises from oscillatory behavior. The direct connection between oscillatory properties and thermodynamic dissipation remains unclear because entropy production is a scalar quantity. For instance, when the entropy production rate changes as a parameter varies, it is challenging to discern whether the change is due to a shift in frequency or amplitude.

To address the aforementioned challenge, we use Koopman mode decomposition~\cite{koopman1931hamiltonian, mezic2013analysis}, a powerful framework for analyzing nonlinear dynamical systems. This method decomposes nonlinear dynamics into a sum of oscillatory modes. The key concept is that nonlinear dynamics are recast as linear evolution in an infinite-dimensional function space governed by a linear operator known as the Koopman generator.  The essential modes that capture this linear evolution can then be identified in a data-driven manner using dynamic mode decomposition (DMD)~\cite{williams2015data, lusch2018deep, otto2019linearly, takeishi2017learning, tu2014on, arbabi2017ergodic, brunton2017chaos, ichinaga2024pydmd, baddoo2023physics}. 
This linearization provides a systematic approach to decomposing a system's behavior into a set of oscillatory modes, thereby making an otherwise difficult-to-interpret system easier to interpret.
This method has also been used to describe noisy nonlinear oscillators~\cite{Thomas2014-aa, Perez-Cervera2021-ku}.

Using Koopman mode decomposition, we establish a general relationship between nonlinear oscillation and the housekeeping entropy production rate in overdamped Langevin systems with general nonlinear forces.
Our central result is the decomposition of the housekeeping entropy production rate into a sum of positive contributions from each Koopman mode. Each mode's contribution is shown to be proportional to the product of its frequency squared and oscillation intensity.
This work is a nonlinear extension of our previous study on linear systems~\cite{sekizawa2024decomposing} and provides a new, interpretable tool for thermodynamic analysis of noise-induced nonlinear oscillatory phenomena.

Furthermore, to demonstrate the utility of our framework in analyzing nonlinear dynamics, we apply it to the noisy FitzHugh–Nagumo model~\cite{fitzhugh1961impulses}, a canonical model for neural excitability. This enables a mode-by-mode analysis of the entropy production rate in the steady state for oscillations undergoing bifurcation and {\color{black} coherent resonance}.
Near the bifurcation threshold, our decomposition revealed that an initially broad spectrum of contributions to the entropy production rate experiences intermittent dropouts as the oscillation fades.
For {\color{black} coherent resonance}, we found that the optimal response to noise was characterized by a broad spectrum of frequency modes, each of which significantly contributed to the total dissipation.
These results provide the first mode-by-mode picture of how thermodynamic dissipation is structured during complex nonlinear events.

\section{Background of stochastic thermodynamics}
\phantomsection
\makeatletter
\def\@currentlabelname{Background of stochastic thermodynamics}
\makeatother
\label{sec:backgruond}
Here we explain the setup of our study. We consider the overdamped multidimensional Langevin equation for the dynamics of the state in $d$-dimensional space, $\bm{x}_t \in \mathbb{R}^{d}$, at time $t$:
\begin{align}
    d\bm{x}_t = \mathit{D}_t\bm{f}_t(\bm{x}_t)dt + \sqrt{2D_t}d\bm{B}_t, \label{eq:Langevin}
\end{align}
where $d\boldsymbol{x}_t$ is the increment of the state; $dt$ is the infinitesimal time interval; $\bm{f}_t(\bm{x}_t)$ is the force at state $\bm{x}_t$ and time $t$, and $\mathit{D}_t$ is a $d \times d$ matrix representing the strength of the noise at time $t$. We assume that $\mathit{D}_t$ is a positive definite matrix, and thus its inverse, $D_t^{-1}$, exists. The term $\sqrt{D_t}$ represents the unique symmetric positive definite square root of $D_t$, which satisfies $D_t = \sqrt{D_t}\sqrt{D_t}^\top$, where the superscript $^\top$ denotes the matrix transpose. The term $d\boldsymbol{B}_t$ denotes an increment of a standard $d$-dimensional Brownian motion, which is a Wiener process satisfying $\mathbb{E}[d\boldsymbol{B}_t]= \boldsymbol{0}$ and $\mathbb{E}[d \boldsymbol{B}_{t}d\boldsymbol{B}_{s}^\top]=\delta(t-s) I dt$, where $\mathbb{E}[\cdot]$ denotes the expected value and $I$ is the identity matrix.

The Langevin equation can be reformulated using the following Fokker--Planck equation:
\begin{align}
    \pdv{p_t(\bm{x})}{t}&=-\nabla \cdot \qty[\bm{\nu}_t(\bm{x})p_t(\bm{x})] \label{eq:Fokker--Planck}\\
    \bm{\nu}_t(\bm{x})&=\mathit{D}_t(\bm{f}_t(\bm{x})-\nabla\ln p_t(\bm{x}))\label{eq:local_mean_velocity}
\end{align}
The Fokker--Planck equation is a deterministic equation describing the temporal evolution of the probability distribution $p_t(\bm{x})$. The velocity field $\bm{\nu}_t(\bm{x})$ is called the local mean velocity. A system is in a steady state when $p_t(\bm{x})$ does not change over time.

The entropy production rate $\sigma_t$ is defined as the following $L^2$ norm of the local mean velocity $\bm{\nu}_t$ with a metric $D_t^{-1}p_t(\bm{x})$ {\color{black}\cite{seifert2025stochastic}}, i.e.,
\begin{align}
    \sigma_t = \langle \bm{\nu}_t^\top \mathit{D}_t^{-1} \bm{\nu}_t \rangle_t, \label{eq:epr_def}
\end{align}
where $\langle \cdots \rangle_t = \int d\bm{x} \ p_t(\bm{x}) \cdots$ denotes the expected value at time $t$. 
We assume that the state variables $\bm{x}_t$ have even parity, meaning they are invariant under time reversal and do not include odd-parity variables such as velocity.
The non-negativity of the entropy production rate, $\sigma_t \ge 0$, is a statement of the second law of thermodynamics~\cite{seifert2025stochastic}.

\subsection{Housekeeping entropy production rate}
\begin{figure}[tbp]
    \centering
    \hspace{-8mm}
    \includegraphics[width=\linewidth]{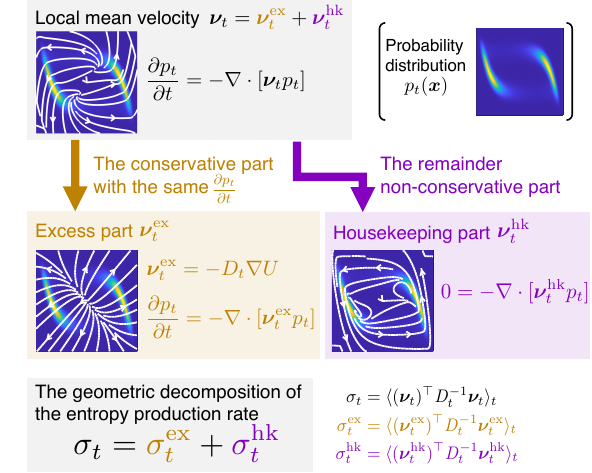}
    \caption{  
    Schematic illustration of the geometric decomposition of the entropy production rate $\sigma_t$ into the housekeeping part $\sigma_t^\mathrm{hk}$ and the excess part $\sigma_t^\mathrm{ex}$ ~\cite{Dechant2022-gt, yoshimura2023housekeeping}. 
    The excess part $\bm{\nu}_t^\mathrm{ex}$ means the velocity field given by the conservative force that provides the same time evolution as the original velocity field $\bm{\nu}_t$. The remainder housekeeping part $\bm{\nu}_t^\mathrm{hk}$ corresponds to the non-conservative force and does not contribute to the time evolution of $p_t(\bm{x})$. 
    The parts of the entropy production rates associated with the respective parts of the local mean velocities are $\sigma_t^\mathrm{ex} = \langle (\bm{\nu}_t^\mathrm{ex})^\top \mathit{D}_t^{-1} \bm{\nu}_t^\mathrm{ex}\rangle_t$ and $\sigma_t^\mathrm{hk} = \langle (\bm{\nu}_t^\mathrm{hk})^\top \mathit{D}_t^{-1} \bm{\nu}_t^\mathrm{hk}\rangle_t$. We use here the noisy FitzHugh--Nagumo model to describe these schematics.
    }
    \label{fig:MN_decomp}
\end{figure}

We consider the housekeeping entropy production rate $\sigma_t^\mathrm{hk}$ introduced by geometric decomposition~\cite{dechant2022geometric} (see also Fig.~\ref{fig:MN_decomp}). This housekeeping quantifies dissipation caused by non-conservative forces. 

To define the housekeeping entropy production rate, we decompose the local mean velocity $\bm{\nu}_t(\boldsymbol{x})$ as $\bm{\nu}_t(\boldsymbol{x}) = \bm{\nu}_t^\mathrm{hk}(\boldsymbol{x}) + \bm{\nu}_t^\mathrm{ex}(\boldsymbol{x})$. Here, $\bm{\nu}_t^\mathrm{ex}(\boldsymbol{x})$ is defined by a potential $U_t(\boldsymbol{x})$, which satisfies the conditions $\bm{\nu}_t^\mathrm{ex}(\boldsymbol{x})= - D_t \nabla U_t(\boldsymbol{x})$ and $\partial p (\bm{x})/\partial t = -\nabla \cdot [\bm{\nu}_t^\mathrm{ex}(\boldsymbol{x}) p_t(\boldsymbol{x})]$. Thus, $\bm{\nu}_t^\mathrm{ex}(\boldsymbol{x})$ means the velocity field given by the conservative force that provides the same time evolution as the original velocity field $\bm{\nu}_t(\boldsymbol{x})$. The remainder $\bm{\nu}_t^\mathrm{hk}(\boldsymbol{x})$ is the contribution that is not given by the conservative force. Because $\partial p_t(\bm{x})/\partial t = -\nabla \cdot [\bm{\nu}_t(\boldsymbol{x}) p_t(\boldsymbol{x})] = -\nabla \cdot [\bm{\nu}_t^\mathrm{ex}(\boldsymbol{x}) p_t(\boldsymbol{x})]$ is satisfied, the term $\bm{\nu}_t^\mathrm{hk}(\boldsymbol{x})$ does not contribute to the time evolution as 
$-\nabla \cdot \qty[ \bm{\nu}_t^\mathrm{hk}(\bm{x}) p_t(\bm{x})] =0$. We note that $\bm{\nu}_t^\mathrm{hk}(\boldsymbol
{x})$ is equivalent to $\bm{\nu}_t(\boldsymbol
{x})$ if $p_t(\bm{x})$ is the steady-state distribution, that is, $\partial p_t (\bm{x})/\partial t =0$. However, $\bm{\nu}_t^\mathrm{hk}(\boldsymbol
{x})$ is not generally given by  the velocity field in the steady state if $p_t(\bm{x})$ is not the steady-state distribution, and it is different from the housekeeping-excess decomposition by Hatano and Sasa~\cite{hatano2001steady}.
The housekeeping entropy production is defined as {\color{black}\cite{dechant2022geometric}}
\begin{align} 
    \sigma^\mathrm{hk}_t = \langle \qty( \bm{\nu}_t^\mathrm{hk} )^\top \mathit{D}_t^{-1} \bm{\nu}_t^\mathrm{hk} \rangle_t, \label{eq:EP_hk} 
\end{align}
which means the dissipation caused by the non-conservative contribution $\bm{\nu}_t^\mathrm{hk}(\boldsymbol{x})$.
The difference between the entropy production rate and the housekeeping entropy production rate is given by the excess entropy production rate $\sigma^\mathrm{ex}_t := \sigma_t - \sigma^\mathrm{hk}_t= \langle \qty( \bm{\nu}_t^\mathrm{ex} )^\top \mathit{D}_t^{-1} \bm{\nu}_t^\mathrm{ex} \rangle_t$, and thus the non-negativity of the excess entropy production rate implies $\sigma^\mathrm{hk}_t \leq \sigma_t$. The equality $\sigma^\mathrm{hk}_t = \sigma_t$ holds in the steady state because $\bm{\nu}_t^\mathrm{ex}(\boldsymbol{x})$ and $\sigma^\mathrm{ex}_t$ become zero when $\partial p_t (\bm{x})/\partial t =0$. We note that the decomposition $\sigma_t =\sigma^\mathrm{ex}_t + \sigma^\mathrm{hk}_t$ is called the geometric decomposition because this decomposition is given by a generalization of the Pythagorean theorem~\cite{dechant2022geometric}.

\section{Koopman mode decomposition of the housekeeping entropy production rate}

\subsection{Koopman mode decomposition of the virtual dynamics driven by $\bm{\nu}_t^\mathrm{hk}$}

To prepare for our main result, we will introduce the Koopman mode decomposition to the following virtual dynamical system:
\begin{align}
    d\bm{x}_s=\bm{\nu}_t^\mathrm{hk}(\bm{x}_s) ds\label{eq:Langevin_hk}, 
\end{align}
where the housekeeping part of the local mean velocity $\bm{\nu}_t^\mathrm{hk} (\bm{x})$ is derived from the original process [Eq.~\ref{eq:Langevin}] to obtain the housekeeping entropy production rate $\sigma_t^\mathrm{hk}$. 
The subscript $s$ stands for the time of the virtual dynamics, whereas $t$ stands for the time of the original Langevin dynamics [Eq.~\ref{eq:Langevin}].
During the virtual deterministic processes, $\bm{\nu}_t^\mathrm{hk} (\bm{x})$ is fixed with respect to changes in $s$. Let $q_s(\bm{x})$ be
the probability distribution at time $s$ in this virtual dynamics. The time evolution of $q_s(\bm{x})$ is given by $\partial q_s(\bm{x})/\partial s = -\nabla \cdot [\bm{\nu}_t^\mathrm{hk} (\bm{x}) q_s(\bm{x})]$. We note that the probability distribution of the original dynamics $p_t(\bm{x})$ becomes the invariant measure of the virtual dynamics. This means that $\partial q_s(\bm{x})/\partial s =0$ if $q_s(\bm{x})$ is the same as $p_t(\bm{x})$.

{\color{black}
Introducing the virtual dynamics is not merely a convenience but is necessary for applying the decomposition to the housekeeping local mean velocity field, which includes both the deterministic drift and the diffusion-induced contribution. 
The oscillations observed in the virtual dynamics therefore incorporate not only the deterministic drift $D_t\bm{f}_t(\bm{x})$, but also the diffusion-induced force $-D_t\nabla \ln p_t(\bm{x})$ from high-density to low-density regions, as defined in Eq.~\ref{eq:local_mean_velocity}. 
In this way, the virtual dynamics retain the effective influence of the stochastic term $\sqrt{2D_t}\,d\bm{B}_t$, which tends to vanish on average in the original Langevin dynamics. Moreover, this step is not an approximation: Eq.~\ref{eq:Langevin_hk} gives an exact deterministic representation of the housekeeping local mean velocity field. 
As discussed further in the Supporting Information, Section \textit{``Physical meaning of frequencies extracted in our methods,''} once diffusion-induced transport is included, the oscillatory structure associated with the local mean velocity field is not always fully characterized by the deterministic drift alone. 
Accordingly, our decomposition captures coherent oscillatory structures shaped jointly by deterministic drift and noise-driven flows. 
}

\begin{figure}[t]
    \centering
    \includegraphics[width=\linewidth]{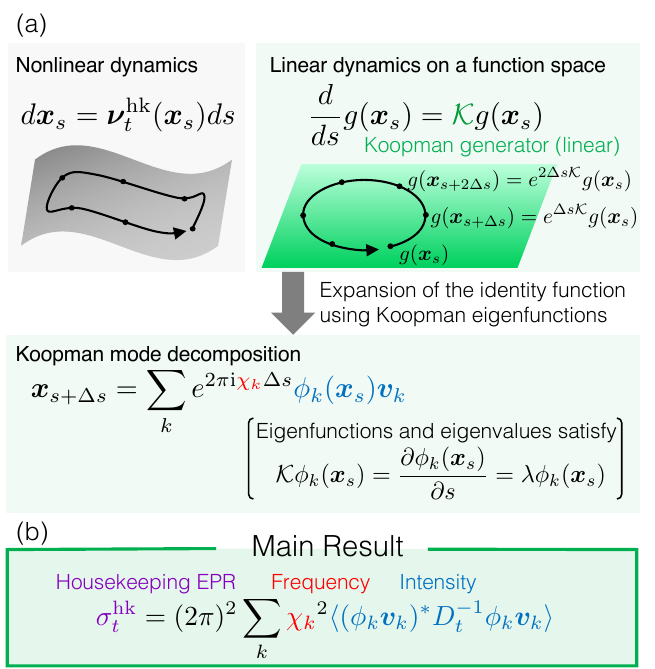}
    \caption{ 
    (a) Koopman mode decomposition. The virtual dynamics in Eq. \ref{eq:Langevin_hk} are decomposed into a sum of the oscillatory modes using the Koopman mode decomposition. 
    The Koopman generator $\mathcal{K}$ transforms a nonlinear dynamical system into the linear dynamics on a function space. 
     Using the eigenvalues $\{\lambda_k\}_{k=1}^r$ and the eigenfunctions $\{\phi_k\}_{k=1}^r$ of the Koopman generator $\mathcal{K}$, the time variation of $\bm{x}_s$ in virtual dynamics can be expressed as a sum of modes. 
    (b) Our main result. The housekeeping entropy production rate is decomposed into a sum of contributions from oscillatory modes. Each mode's contribution is the product of the square of its frequency and its oscillation intensity.         
    }\label{fig:Koopman}
\end{figure}

The nonlinear dynamics in Eq.~\ref{eq:Langevin_hk} can be represented as linear dynamics by extracting a finite number of modes through Koopman mode decomposition~\cite{koopman1931hamiltonian, mezic2013analysis} (see Fig. \ref{fig:Koopman}a). As detailed in Supporting Information (Section \textit{“Koopman mode decomposition of virtual dynamics given by $\bm{\nu}_t^\mathrm{hk}$”}), the Koopman generator $\mathcal{K}$ is defined as a linear operator that maps a function $g: \mathbb{R}^d \to \mathbb{C}$ to another function $\mathcal{K}g:= \nabla g^\top \bm{\nu}_t^\mathrm{hk}$. 
This linear operator satisfies $\mathcal{K}g(\bm{x}_s) = (d/ds)g(\bm{x}_s)$, describing the linear time evolution of the observable $g(\bm{x}_s)$. Intuitively, the Koopman generator  transforms a nonlinear dynamical system into a linear dynamical system on a function space, which is driven by the linear operator $\mathcal{K}$.

To describe the nonlinear dynamics by exploiting this linearization, we consider an  expansion of the identity function $\mathrm{Id}(\bm{x})=\bm{x}$ using the eigenvalues $\{\lambda_k\}_{k=1}^r$ and the eigenfunctions $\{\phi_k\}_{k=1}^r$ of the Koopman generator $\mathcal{K}$ satisfying $\mathcal{K}\phi_k(\bm{x}) =\lambda_k \phi_k(\bm{x})$. The scalar $r$ is the number of modes. 
In this study, we assume that the Koopman generator can be accurately approximated by a finite-dimensional linear operator.
In the virtual dynamics in Eq. \ref{eq:Langevin_hk}, the Koopman generator $\mathcal{K}$ is skew-adjoint  {\color{black}with respect to the $L^2$ inner product weighted by $p_t(\bm{x})$} (see Supporting Information, Section \textit{“Skew-adjointness and diagonalizability of the Koopman generator”}) and is therefore diagonalizable under this finite-dimensional approximation. 
Then, the identity function is given by $\mathrm{Id}(\bm{x}_s) = \sum_k^r \phi_k(\bm{x}_s) \bm{v}_k$ with weights $\bm{v}_k$ called Koopman modes, and 
the time variation of $\bm{x}_s$ on the virtual dynamics can be written as
\begin{align}
    \bm{x}_{s+\Delta s} &= \sum_{{\color{black} k=1}}^r e^{\lambda_k \Delta s} \phi_k(\bm{x}_s) \bm{v}_k = \sum_{{\color{black} k=1}}^r e^{2\pi\chi_k \mathrm{i}\Delta s} \phi_k(\bm{x}_s) \bm{v}_k \label{eq:oscillation_x}.
\end{align}
This decomposition is called the Koopman mode decomposition. Here, $\chi_k$ is the frequency defined as \begin{align}
    \chi_k = \left|\lambda_k/(2\pi \mathrm{i}) \right|, \label{eq:def_chi}
\end{align} where $\mathrm{i}$ stands for  the imaginary unit. As derived in Supporting Information (Section \textit{“Derivation of the main result”}), the eigenvalue $\lambda_k$ is purely imaginary, and hence, $\chi_k$ is a real number. This means that the time variation of $\bm{x}_s$ in Eq. \ref{eq:oscillation_x} can be expressed as the sum of the oscillatory modes.

We also introduce the intensities of the oscillatory modes. When the eigenvalues are not degenerate, the intensity of the $k$-th oscillatory mode is \begin{align}
    J_k = \langle (\phi_k\bm{v}_k)^*D_t^{-1}(\phi_k\bm{v}_k)\rangle_t. \label{eq:def_normalized_intensity}
\end{align}
The symbol $^*$ stands for conjugate transpose. This quantity is the $L^2$ norm of the $k$-th mode $\phi_k \bm{v}_k$ in Eq.~\ref{eq:oscillation_x} under the metric $D_t^{-1}p_t(\bm{x})$, representing the strength of $k$-th oscillatory mode.

\subsection{The main result}
Our main result is a decomposition of the housekeeping entropy production rate into independent positive contributions from each oscillatory mode: 
\begin{align}
    \sigma_t^\mathrm{hk} &= \sum_k \sigma_t^{\mathrm{hk}, (k)} \nonumber\\
    \sigma_t^{\mathrm{hk}, (k)} &=(2\pi)^2\chi_k^2 J_k. \label{eq:decomposition_EP}
\end{align}
(For the derivation of this decomposition, see Supporting Information, Section \textit{“Derivation of the main result”}). 
The decomposition means that the contribution of each oscillatory mode to the housekeeping entropy production rate is the product of its frequency squared $\chi_k^2$ and its intensity $J_k$ (see also Fig. \ref{fig:Koopman}b). In other words, modes with higher frequencies and greater intensities contribute more to the housekeeping entropy production rate.

When the eigenvalues are degenerate, the decomposition becomes
\begin{align}
    \sigma_t^\mathrm{hk} &= \sum_{k'} (2\pi)^2\chi_{k'}^2 \left\langle \qty(\sum_{l\in C_{k'}}\phi_l\bm{v}_l)^*  {D_t^{-1}}\qty(\sum_{m\in C_{k'}}\phi_m\bm{v}_m)\right\rangle_t ,\label{eq:decomposition_EP_degenerate}
\end{align}
\textcolor{black}{where $C_{k'}:=\{l \mid \lambda_l = 2\pi \chi_{k'} \mathrm{i}\}$ is the set of indices corresponding to the degenerate eigenvalue $2\pi i\chi_k$. Here, the index $k'$ is defined such that each $\chi_{k'}$ is distinct. The summation is taken over only those $k'$ for which $\chi_{k'}$ has different values, ensuring that no $k'$ with the same value of $\chi_{k'}$ is selected more than once.}

We note that our decomposition in Eq.~\ref{eq:decomposition_EP} is an extension of our previous result for linear Langevin systems \cite{sekizawa2024decomposing}. We can derive the previous result in Ref.~\cite{sekizawa2024decomposing} from Eq.~\ref{eq:decomposition_EP} (see Supporting Information, Section \textit{“Linear Langevin dynamics”}).

We also note that although our method is based on the eigenvalues of the Koopman generator, it differs from previous approaches that rely on the eigenvalues of transition rate matrices in discrete-state Markov processes
~\cite{Uhl2019-ch, kolchinsky2024thermodynamic,xu2025thermodynamic, Oberreiter2022-nx, Del-Junco2020-px,barato2017coherence, ohga2023thermodynamic}.
In our framework, we consider a virtual deterministic dynamical system, meaning that all Koopman eigenvalues are purely imaginary. Instead, we would like to point out that this decomposition is similar to the cycle decomposition of housekeeping entropy production rates~\cite{yoshimura2023housekeeping}, which is based on Schnakenberg's network theory~\cite{schnakenberg1976network}.

{\color{black}
The decomposition also implies a constraint relating the oscillation frequency, oscillation intensity, and thermodynamic cost. 
Since all contributions in Eq.~\ref{eq:decomposition_EP} are non-negative, each oscillatory mode must satisfy
\begin{align}
J_k
\le
\frac{1}{(2\pi \chi_k)^2}\,
\sigma_t^{\mathrm{hk}} \le  \frac{1}{(2\pi \chi_k)^2}\,
\sigma_t. \label{eq:bound}
\end{align}
for $\chi_k \neq 0$.
This inequality shows that, under a finite thermodynamic cost $\sigma_t$, the achievable oscillation intensity $J_k$ is bounded. 
In particular, the bound decreases with increasing frequency, indicating that sustaining strong high-frequency oscillations requires a larger entropy production rate.
}

\section{Applications to the noisy FitzHugh--Nagumo model}
\begin{figure*}[htbp]
    \centering
    \includegraphics[width=\linewidth]{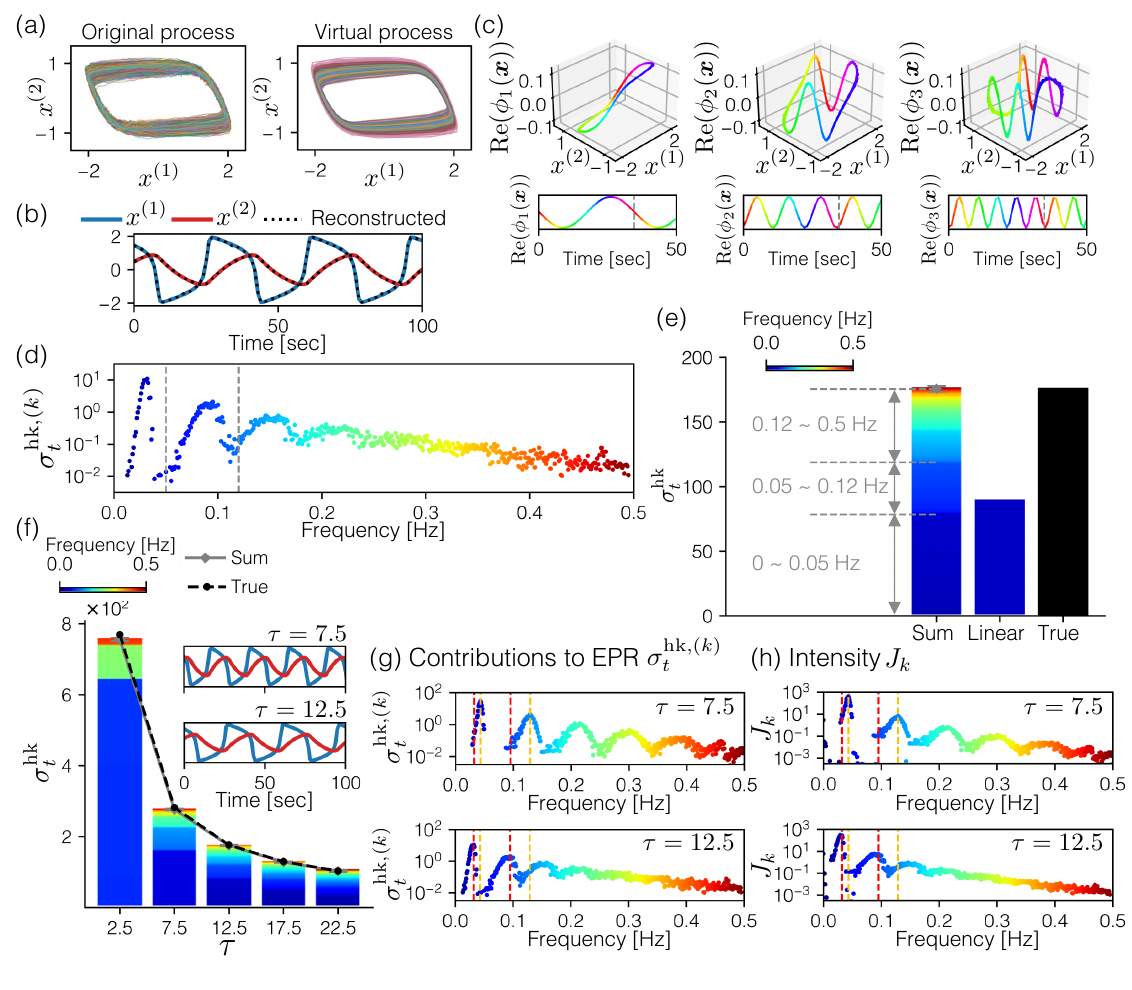}
    \vspace{-8mm}
    \caption{
    An application example to the noisy FitzHugh--Nagumo model. 
    (a) Examples of trajectories that follow the original Langevin process in Eq. \ref{eq:Langevin_FN} and the virtual deterministic process in Eq. \ref{eq:Langevin_hk}.
    (b) An example of a trajectory following the virtual dynamics of the noisy FitzHugh--Nagumo model driven by the housekeeping part of the local mean velocity [Eq.~\ref{eq:Langevin_hk}] (colored lines) and the dynamics reconstructed from the Koopman mode decomposition [Eq.~\ref{eq:oscillation_x}] (black dotted line). 
    (c) Koopman eigenfunctions $\phi_k(\bm{x})$ along an example trajectory. 
    Top: The value of $\text{Re}(\phi_k(\bm{x}))$ at each point along the trajectory. The color represents the time $s$ modulo the period of the slowest oscillation $1/\chi_1$, and is consistent with the color used in the bottom panel. 
    Bottom: The temporal evolution of $\text{Re}(\phi_k(\bm{x}))$ along the trajectory.
    (d) The contribution of each oscillatory mode to the housekeeping entropy production rate $\sigma_t^{\mathrm{hk}, (k)}$. Each dot represents the contribution $\sigma_t^{\mathrm{hk}, (k)}$. The vertical dashed lines at 0.05 Hz and 0.12 Hz are provided to facilitate comparison with (e).
    Top: Result from a single trajectory. 
    Bottom: Results from 1000 trajectories, computed using a moving window over frequency.
    (e) The sum of the contributions from (d) almost equals the total value of the housekeeping entropy production rate. 
    Left: a stacked bar plot of $\sigma_t^{\mathrm{hk}, (k)}$. 
    The colors represent the frequencies of the oscillatory modes. 
    The error bars indicate 95\% confidence intervals of the sum of the contributions.     
    Middle: a stacked bar plot of $\sigma_t^{\mathrm{hk}, (k)}$ under the assumption of the linear dynamics, calculated using the methods in Ref.~\cite{sekizawa2024decomposing}. 
    The colors represent the frequencies of the oscillatory modes. 
    Right: the true housekeeping entropy production rate. 
    (f) Stacked bar plots showing our decomposition for different values of the time constant $\tau$ in Eq.~\ref{eq:Langevin_FN}. 
    The stacked bar plots show the sum of the contributions from the oscillatory modes. 
    The colors represent the frequencies of the oscillatory modes. 
    The gray line indicates the sum of the contributions from the oscillatory modes, with error bars representing 95\% confidence intervals. 
    The black dashed line shows the true housekeeping entropy production rate. 
    The insets represent examples of the trajectories. 
    (g) The contribution of each oscillatory mode to the housekeeping entropy production rate $\sigma_t^{\mathrm{hk}, (k)}$ for different time constants $\tau$, computed using a moving window over frequency. The vertical dashed lines make it easier to compare peak positions; the red and orange lines respectively indicate the peaks for $\tau = 7.5$ and $\tau = 12.5$.
    (h) The intensities of the oscillatory modes. The vertical dashed lines also make it easier to compare peak positions; the red and orange lines respectively indicate the peaks for $\tau = 7.5$ and $\tau = 12.5$.
    }
    \label{fig:FN_demo}
\end{figure*}

\phantomsection
\makeatletter
\def\@currentlabelname{Applications to the noisy FitzHugh--Nagumo model}
\makeatother
\label{sec:fitzhugh}
We demonstrate our decomposition using the FitzHugh–Nagumo model~\cite{fitzhugh1961impulses} in the presence of noise.
This analysis has two main objectives: 
(i) to facilitate an intuitive understanding of our decomposition in a nonlinear setting, and
(ii) to demonstrate its utility in investigating how nonlinear oscillatory phenomena generate the housekeeping entropy production rate.
To achieve the first objective, Fig.~\ref{fig:FN_demo} illustrates how our decomposition represents the housekeeping entropy production rate through oscillatory modes.
For the second objective, in Figs.~\ref{fig:FN_I} and \ref{fig:FN_T}, we apply the decomposition to bifurcations and {\color{black} coherent resonance}.
Throughout, we analyze the steady state, so that the housekeeping entropy production rate coincides with the steady-state entropy production rate.
To estimate the Koopman modes and eigenfunctions used in our decomposition, we employed dynamic mode decomposition (DMD)~\cite{tu2014on, arbabi2017ergodic, brunton2017chaos, baddoo2023physics, ichinaga2024pydmd}. The analysis methods are detailed in \textit{``\nameref{sec:Methods}''}.
{\color{black}We note that the extension to non-steady-state settings remains feasible in a low-dimensional relaxation process; a representative example and the corresponding numerical procedures are given in Supporting Information, in the sections ``Application of our decomposition to non-steady-state dynamics'' and ``Supplementary Methods''

}.

The noisy FitzHugh--Nagumo model is given by the following Langevin equation: \begin{align}
    d\mqty(x^{(1)}_t\\ x^{(2)}_t)=\mqty(x^{(1)}_t-\frac{(x^{(1)}_t)^3}{3}-x^{(2)}_t+I\\ \frac{1}{\tau}(x^{(1)}_t+a-bx^{(2)}_t))dt+\sqrt{2T}d\bm{B}_t \label{eq:Langevin_FN}
\end{align}
The superscripts $^{(i)}$ for $i\in \{1,2\}$ represent indices of the dimensions of the state vector $\bm{x}_t = (x^{(1)}_t, x^{(2)}_t)^{\top}$.
The noisy FitzHugh--Nagumo model is a neuronal model~\cite{fitzhugh1961impulses} with a membrane potential $x^{(1)}_t\in \mathbb{R}$ and a recovery variable  $x^{(2)}_t\in \mathbb{R}$ of a neuron. 
The parameters $a\in \mathbb{R}$ and $b\in \mathbb{R}$ reflect properties of the neuron. 
$I\in \mathbb{R}$ is an input to the neuron. 
$\tau {\color{black}\in \mathbb{R}_{>0}}$ is the time-constant of the recovery variable. 
$T {\color{black}\in \mathbb{R}_{>0}}$ is the intensity of the noise. 
For Fig. \ref{fig:FN_demo}, we chose the following values: $a=0, b=0.5, I=0, T=10^{-3}$. The parameter $\tau=12.5$ is used for Figs. \ref{fig:FN_demo}a-e, and the parameter $\tau$ varies for Figs. \ref{fig:FN_demo}f-h.
With these parameter values, the dynamics exhibit an oscillatory pattern (see Fig. \ref{fig:FN_demo}a, left).

\subsection{Demonstration of our decomposition}
In this section, we illustrate our decomposition.
To prepare for this, we numerically calculated the housekeeping part of the local mean velocity $\bm{\nu}_t^\mathrm{hk}(\boldsymbol{x})$ and simulated time-series data that follow the virtual dynamics driven by $\bm{\nu}_t^\mathrm{hk}(\boldsymbol{x})$, as defined in Eq.~\ref{eq:Langevin_hk}, which mimics the original Langevin dynamics in Eq.~\ref{eq:Langevin_FN} (see Fig.~\ref{fig:FN_demo}a, right, and Fig.~\ref{fig:FN_demo}b, colored lines).

We applied the Koopman mode decomposition to each simulated trajectory to characterize how the virtual dynamics can be expressed in terms of oscillatory modes.
From this analysis, we obtained the eigenvalues $\{\lambda_k\}_{k=1}^r$, eigenfunctions $\{\phi_k\}_{k=1}^r$, and Koopman modes $\{\bm{v}_k\}_{k=1}^r$ associated with the virtual dynamics.
In the example trajectory shown in Fig.~\ref{fig:FN_demo}b, {\color{black}the trajectory was well reconstructed using $27$ modes (black dashed line) \color{black}(see also Materials and Methods)}. 
The real parts of the eigenfunctions, ${\rm Re}(\phi_k(\bm{x}))$, exhibit oscillatory variations as $\bm{x}$ evolves along the trajectory with non-uniform speed, returning to their original values after one cycle (see Fig.~\ref{fig:FN_demo}c, top).
When plotted against time, ${\rm Re}(\phi_k(\bm{x}))$ displays sinusoidal waveforms (see Fig.~\ref{fig:FN_demo}c, bottom).
These results indicate that the dynamics of the virtual system can be interpreted as a superposition of oscillatory modes [Eq.~\ref{eq:oscillation_x}], from which the time series can be faithfully reconstructed (see Fig.~\ref{fig:FN_demo}b, black dashed line).
{\color{black}
In the calculation of our decomposition, the expectations are evaluated by Monte Carlo sampling over many trajectories of this virtual dynamics.
Because the true underlying Langevin dynamics is known in the present setting, any discrepancy caused by an insufficient number of sampled trajectories can in practice be checked and reduced by increasing the sample size until convergence is reached while monitoring the confidence intervals.
See the Supporting Information, ``Finite sampling effects in low noise metastable systems,'' for details.

}

From our decomposition in Eq.~\ref{eq:decomposition_EP}, we obtain the contributions of the oscillatory modes to the housekeeping entropy production rate $\sigma_t^{\mathrm{hk},(k)}$ (see Figs. \ref{fig:FN_demo}d-e). We observe large contributions from frequencies around $0.03$ Hz, $0.08$ Hz, and so on (see Fig. \ref{fig:FN_demo}d). The sum of these contributions recovers the total housekeeping entropy production rate $\sigma_t^\mathrm{hk}$ (see Fig. \ref{fig:FN_demo}e, left). This result is consistent with the true housekeeping entropy production rate calculated numerically using the methods in Ref. \cite{gingrich2017inferring} (see Fig. \ref{fig:FN_demo}e, right), indicating the validity of our decomposition.

In contrast, when applying the method in Ref.~\cite{sekizawa2024decomposing}, which assumes linear Langevin dynamics, the sum of the contributions of modes does not recover the true housekeeping entropy production rate $\sigma_t^\mathrm{hk}$ (see Fig. \ref{fig:FN_demo}e, middle). This result demonstrates the necessity of handling nonlinear dynamics in our decomposition to accurately capture the contributions of oscillatory modes in nonlinear systems.

Next, we analyzed how the decomposition varies with the time constant $\tau$, to demonstrate how our decomposition reveals frequency-dependent features of thermodynamic dissipation that are not accessible from the total entropy production rate alone.
As $\tau$ increases, the virtual dynamics exhibit oscillations with lower frequencies (see Fig.~\ref{fig:FN_demo}f, insets), and the total housekeeping entropy production rate $\sigma_t^\mathrm{hk}$ without decomposition decreases correspondingly (see Fig.~\ref{fig:FN_demo}f, black dashed line).
Our decomposition allows us to interpret this decrease in terms of frequency-resolved contributions of oscillatory modes.
As $\tau$ increases, modes with smaller frequencies become dominant contributors to $\sigma_t^\mathrm{hk}$ (see Figs.~\ref{fig:FN_demo}f,g), and the overall dissipation decreases because each mode’s contribution $\sigma_t^{\mathrm{hk},(k)}$ scales with the square of its frequency [Eq.~\ref{eq:decomposition_EP}].
This correspondence between frequency and dissipation is consistent with the theoretical structure of our decomposition, which attributes higher energetic costs to faster oscillations.
Despite the frequency shift between $\tau = 7.5$ and $\tau = 12.5$, the peak intensities remain nearly constant (see Fig.~\ref{fig:FN_demo}h), indicating that the reduction in $\sigma_t^\mathrm{hk}$ primarily arises from lower oscillation frequencies rather than changes in their amplitudes.

\subsection{Application of our decomposition across bifurcation regimes}

\begin{figure*}[ht]
    \centering
    \includegraphics[width=\linewidth]{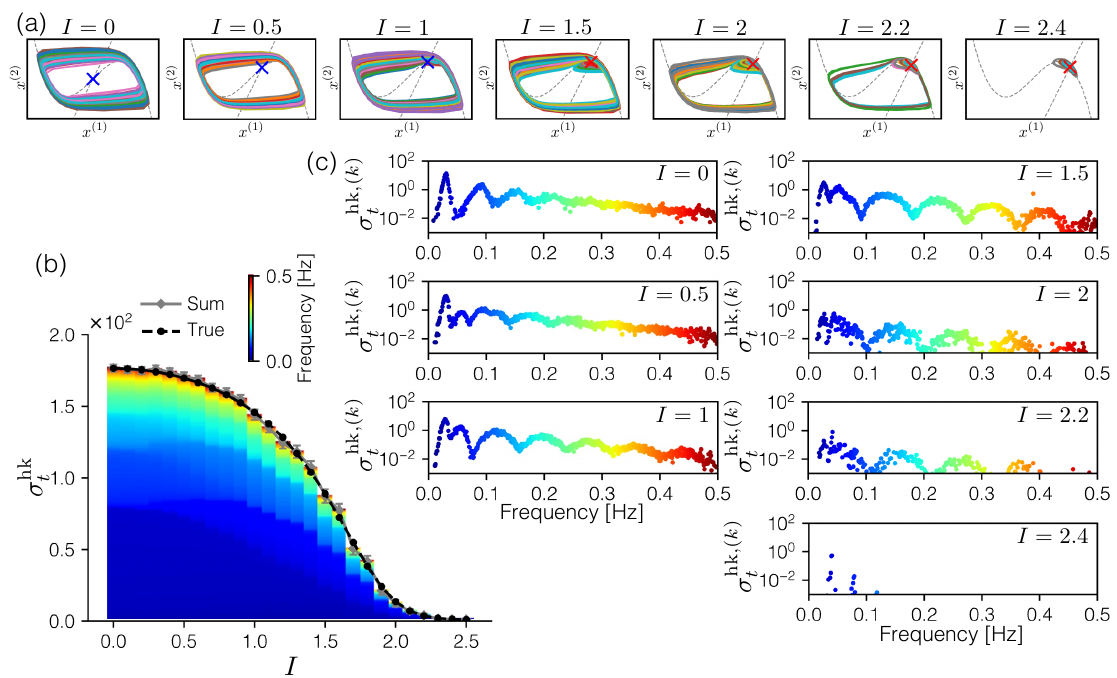}
    \vspace{-5mm}
    \caption{
    Our decomposition enables us to understand how the entropy production rate depends on the parameter  $I$ near the bifurcation point of the noisy FitzHugh–Nagumo model.
    (a) Examples of trajectories of the virtual dynamics in Eq.~\ref{eq:Langevin_hk} for the noisy FitzHugh--Nagumo model for different values of $I$. 
    The black dashed line represents the nullclines of the noisy FitzHugh--Nagumo model, which were calculated from the original Langevin dynamics in Eq.~\ref{eq:Langevin_FN} by ignoring the noise term. 
    The blue and red crosses represent the unstable and stable fixed points, respectively.
    For $I < 1.5$, the trajectory forms a large loop around the unstable fixed point. However, near $I = 1.5$, a stable fixed point emerges, and the trajectory transitions to a smaller loop around this stable point. 
    (b) The parameter-dependent behavior of the housekeeping entropy production rate and its decomposition. 
    The stacked bar plot shows the sum of the contributions from the oscillatory modes. 
    The colors represent the frequencies of the oscillatory modes. 
    The gray line indicates the sum of the contributions from the oscillatory modes, with error bars representing 95\% confidence intervals. 
    The black dashed line shows the true housekeeping entropy production rate. 
    As the trajectory transitions from the large loop to the small loop, the housekeeping entropy production rate $\sigma_t^\mathrm{hk}$ significantly decreases. 
    (c) The contribution of each oscillatory mode to the housekeeping entropy production rate $\sigma_t^{\mathrm{hk}, (k)}$ for different input  values of $I$. For $I<2.4$, a variety of frequencies contribute to the housekeeping entropy production rate $\sigma_t^\mathrm{hk}$. As $I$ approaches $2.4$, the contributions from frequencies undergo intermittent dropout. At $I=2.4$, almost a single frequency predominantly contributes to the housekeeping entropy production rate $\sigma_t^\mathrm{hk}$.
    }
    \label{fig:FN_I}
\end{figure*}

Building on the demonstration in the previous sections, we next apply our decomposition to investigate how the housekeeping entropy production rate arises near the bifurcation points of the noisy FitzHugh–Nagumo model, thereby demonstrating its utility for analyzing thermodynamic dissipation in nonlinear phenomena.
The FitzHugh–Nagumo model is known to exhibit bifurcations in which the qualitative behavior of the system changes depending on parameters such as $I$ and $b$~\cite{fitzhugh1961impulses, murray2007mathematical}.
Fig.~\ref{fig:FN_I}a shows the trajectories of the virtual dynamics in Eq.~\ref{eq:Langevin_hk} for various values of $I$.
When $I$ is small, the trajectory forms a large loop, whereas, as $I$ increases, it transitions to a smaller loop.
In the noise-free limit, this transition corresponds to a Hopf bifurcation~\cite{strogatz2001nonlinear}, beyond which the small loop disappears.
However, in the noisy FitzHugh–Nagumo model, the small loop persists even after the bifurcation due to the presence of weak noise.

Prior to analyzing the housekeeping entropy production rate, we first review how bifurcations emerge in the noisy FitzHugh--Nagumo model by examining the fixed points \cite{murray2007mathematical}. 
Figure~\ref{fig:FN_I}a shows the fixed points calculated by ignoring the noise term  in the original Langevin system [Eq. \ref{eq:Langevin_FN}].  
Unstable fixed points are indicated by blue crosses, and stable fixed points are indicated by red crosses.
As the parameter $I$ increases, an unstable fixed point becomes stable around $I = 1.5$ (see Fig.~\ref{fig:FN_I}a). Trajectories that initially form large loops with small $I$ values begin shifting toward smaller loops as they approach the Hopf bifurcation. As $I$ increases further, the small loops gradually become dominant. This gradual transition, rather than an abrupt change, reflects the influence of diffusion due to the noise term.

Before applying our decomposition, we examined the total housekeeping entropy production rate and found that it exhibits a gradual decrease near the bifurcation points (see Fig.~\ref{fig:FN_I}b).
As the input parameter $I$ increases toward $I = 1.5$, where an unstable fixed point becomes stable, the total housekeeping entropy production rate decreases sharply, corresponding to the transition from large loops to small loops (see Fig.~\ref{fig:FN_I}a).
With further increases in $I$, the total housekeeping entropy production rate approaches zero around $I = 2.4$, reflecting the dominance of small-loop trajectories around stable fixed points.
Although this decrease in the total housekeeping entropy production rate captures the overall effect of the bifurcation, it offers no insight into how individual oscillatory modes contribute to this change.

Qualitative changes in the underlying contributions to the housekeeping entropy production rate from oscillatory modes are revealed only through our decomposition.
When the input parameter $I$ is small (around $I \simeq 0$), oscillatory modes spanning a broad range of frequencies significantly contribute to the entropy production rate (see Fig.~\ref{fig:FN_I}c). 
However, as $I$ increases around the bifurcation point $I \simeq 1.5$, these contributions exhibit intermittent dropouts in specific frequency bands. 
This behavior is consistent with the observation that some trajectories become trapped in small loops around stable fixed points. These small loops are associated with low-intensity oscillations that negligibly contribute to the entropy production rate in our decomposition [Eq.~\ref{eq:decomposition_EP}]. 
At around $I = 2.4$, one frequency dominates, exceeding all others by nearly two orders of magnitude. 
This occurs because, in the weak-noise limit, the system behaves like a two-dimensional linear dynamical system around a stable fixed point. In such cases, only one pair of complex conjugate eigenvalues remains, resulting in a single dominant oscillatory mode.
Together, these results demonstrate that our decomposition provides insights unavailable from the total entropy production rate alone, offering a quantitative means to understand how bifurcations in nonlinear systems shape their thermodynamic dissipation.

{\color{black}
\subsection{Physical interpretation of the oscillatory frequencies extracted by our decomposition}
\begin{figure}[t]
    \centering
    \includegraphics[width=\linewidth]{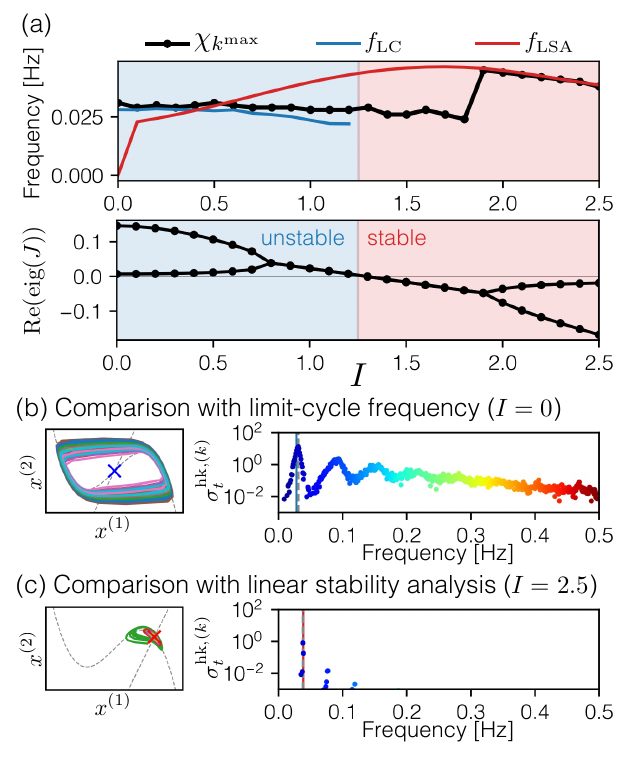}
    \vspace{-5mm}
    \caption{\color{black} 
    Comparison between the oscillatory frequency extracted by our decomposition and 
    (i) the limit cycle frequency $f_{\mathrm{LC}}$ and 
    (ii) the frequency $f_{\mathrm{LSA}}$ predicted by linear stability analysis of the local mean velocity field $\bm{\nu}_t^{\mathrm{hk}}$.
    (a) Top: Frequency $\chi_{k^\text{max}} (>0)$ contributing most strongly to the housekeeping entropy production rate (black dots), plotted as a function of the input current $I$.
    The blue curve indicates the limit cycle frequency $f_{\mathrm{LC}}$ when it exists.
    The red curve shows the frequency $f_{\mathrm{LSA}}$ obtained from the imaginary part of the eigenvalues of the Jacobian of $\bm{\nu}_t^{\mathrm{hk}}$ evaluated at its fixed point satisfying $\bm{\nu}_t^{\mathrm{hk}}=0$.
    Bottom: Real parts of the two eigenvalues of the Jacobian of the deterministic drift, $J = \partial (D_t \bm{f}_t) / \partial \bm{x}|_{\boldsymbol{x}=\boldsymbol{x}_{\rm dfix}}$, at the deterministic fixed point $\boldsymbol{x}_{\rm dfix}$ (i.e., $D_t \bm{f}_t (\boldsymbol{x}_{\rm dfix})=\boldsymbol{0}$), indicating unstable (blue shaded) and stable (red shaded) regimes.
    (b,c) Examples in the limit cycle regime ($I=0$) and the stable fixed point regime ($I=2.5$), respectively.
    Left: Examples of trajectories of the virtual dynamics.
    The blue and red crosses indicate the unstable and stable fixed points, respectively.
    Right: Contributions of oscillatory modes to the housekeeping entropy production rate.
    The gray vertical line indicates $\chi_{k^\text{max}}$, while the blue and red vertical lines indicate $f_{\mathrm{LC}}$ and $f_{\mathrm{LSA}}$, respectively.
    }
    \label{fig:oscillation_main}
\end{figure}

Having analyzed how the entropy production rate is distributed among oscillatory modes across the bifurcation, we next interpret the dominant frequencies extracted by our decomposition by comparing them with two characteristic frequencies in this system: 
(i) the limit cycle frequency $f_{\mathrm{LC}}$ and (ii) the frequency $f_{\mathrm{LSA}}$ predicted by linear stability analysis (see Fig.~\ref{fig:oscillation_main}). 
The relation to (iii) the frequencies associated with the stochastic generator spectrum is also discussed in Supporting Information, Section \textit{“Physical meaning of frequencies extracted in our methods”}. 

For this analysis, the key theoretical point is that the oscillatory frequencies extracted by our decomposition are naturally associated with the local mean velocity field $\bm{\nu}_t^{\mathrm{hk}}$, which is determined by both the deterministic drift $D_t\bm{f}_t(\bm{x})$ and the diffusion-induced contribution $-D_t \nabla \ln p_t(\bm{x})$.
A more detailed discussion of these two contributions is given in Supporting Information, Section \textit{``Physical meaning of frequencies extracted in our methods''}. 
Here, we illustrate this interpretation through the representative noisy FitzHugh--Nagumo example in Fig.~\ref{fig:oscillation_main}, showing how the dominant frequency in our decomposition is related to the limit-cycle frequency and to the frequency predicted by linear stability analysis of $\bm{\nu}_t^{\mathrm{hk}}$.

To compare the dominant frequency extracted by our decomposition with these reference frequencies, we introduce the quantities used in Fig.~\ref{fig:oscillation_main}.
Here, \(f_{\mathrm{LC}}\) denotes the limit-cycle frequency, defined as the reciprocal of the mean time required for the phase of the Langevin dynamics to advance by one cycle; its detailed definition and numerical evaluation are described in Supporting Information, Section \textit{“Methods for Supporting Information”}.
We next define the frequency predicted by linear stability analysis and the dominant frequency extracted by our decomposition. 
Let $\bm{x}_{\mathrm{fix}}$ be the fixed point satisfying $\bm{\nu}_t^{\mathrm{hk}}(\bm{x}_{\mathrm{fix}})=0$, and let $\lambda_{\mathrm{LSA}}$ denote the eigenvalue of the Jacobian $\partial \bm{\nu}_t^{\mathrm{hk}}/\partial \bm{x}|_{\bm{x}=\bm{x}_{\mathrm{fix}}}.$
In the present two-dimensional noisy FitzHugh--Nagumo setting, the eigenvalues form either a complex conjugate pair or a pair of real eigenvalues, so $\lambda_{\mathrm{LSA}}$ is uniquely defined whenever the eigenvalues are complex.
We then define $f_{\mathrm{LSA}}
    =\left|\mathrm{Im}\!\left(\lambda_{\mathrm{LSA}}\right)\right|/(2\pi)$,
which represents the oscillation frequency predicted by linear stability analysis of $\bm{\nu}_t^{\mathrm{hk}}$ around its fixed point. 
We also introduce $ k^\text{max}=\argmax_k \ \sigma_t^{\mathrm{hk}, (k)}$, where $\chi_{k^\text{max}}$ is the frequency defined by \eqref{eq:def_chi} with \(k=k^\text{max}\), namely, the frequency contributing most strongly to the housekeeping entropy production rate in our decomposition.
Here, $k^\text{max}$ does not depend on $t$ because we analyze the steady state.
This figure uses the same parameters as Fig.~\ref{fig:FN_I}.

With this analysis, we found that in the noisy FitzHugh--Nagumo system studied here, the dominant frequency \(\chi_{k^\text{max}}\) extracted by our decomposition roughly coincides with the limit-cycle frequency \(f_{\mathrm{LC}}\) in the unstable regime and with the frequency \(f_{\mathrm{LSA}}\) predicted by linear stability analysis in the stable regime.
In Fig.~\ref{fig:oscillation_main}a (bottom), the real parts of the eigenvalues of the Jacobian $J=\partial (D_t \bm{f}_t)/\partial \bm{x}|_{\bm{x}=\bm{x}_{\mathrm{dfix}}}$ classify the deterministic fixed point $\bm{x}_{\mathrm{dfix}}$ (i.e., $D_t \bm{f}_t (\boldsymbol{x}_{\rm dfix})=\boldsymbol{0}$) into an unstable regime at small $I$ (blue shaded region) and a stable regime at large $I$ (red shaded region).
In the present system, these correspond to a limit cycle regime and a regime in which the deterministic dynamics remain near a stable fixed point, respectively.
Figure~\ref{fig:oscillation_main}a (top) shows, for each value of the input current $I$, the frequency $\chi_{k^\text{max}}$ together with the limit cycle frequency $f_{\mathrm{LC}}$ when it exists (blue curve) and the frequency $f_{\mathrm{LSA}}$ predicted from the fixed-point analysis of $\bm{\nu}_t^{\mathrm{hk}}$ (red curve).
As shown there, $\chi_{k^\text{max}}$ roughly coincides with $f_{\mathrm{LC}}$ throughout the unstable regime, whereas it roughly coincides with $f_{\mathrm{LSA}}$ throughout the stable regime.
This correspondence is illustrated by representative examples: at $I=0$, where the dynamics exhibit limit-cycle-like behavior, $\chi_{k^\text{max}}$ agrees with $f_{\mathrm{LC}}$ (Fig.~\ref{fig:oscillation_main}b), whereas at $I=2.5$, where the dynamics remain near a stable fixed point, $\chi_{k^\text{max}}$ agrees with $f_{\mathrm{LSA}}$ (Fig.~\ref{fig:oscillation_main}c).

Taken together, these results provide a physical interpretation of the oscillatory frequencies identified by our decomposition.
In the present example, these frequencies track the relevant oscillatory motion in each regime: the limit cycle frequency in the unstable regime and the fixed-point oscillation frequency of $\bm{\nu}_t^{\mathrm{hk}}$ in the stable regime.

}

\subsection{Application of our decomposition to {\color{black}coherent resonance}}

\begin{figure*}[t]
    \centering
    \includegraphics[width=\linewidth]{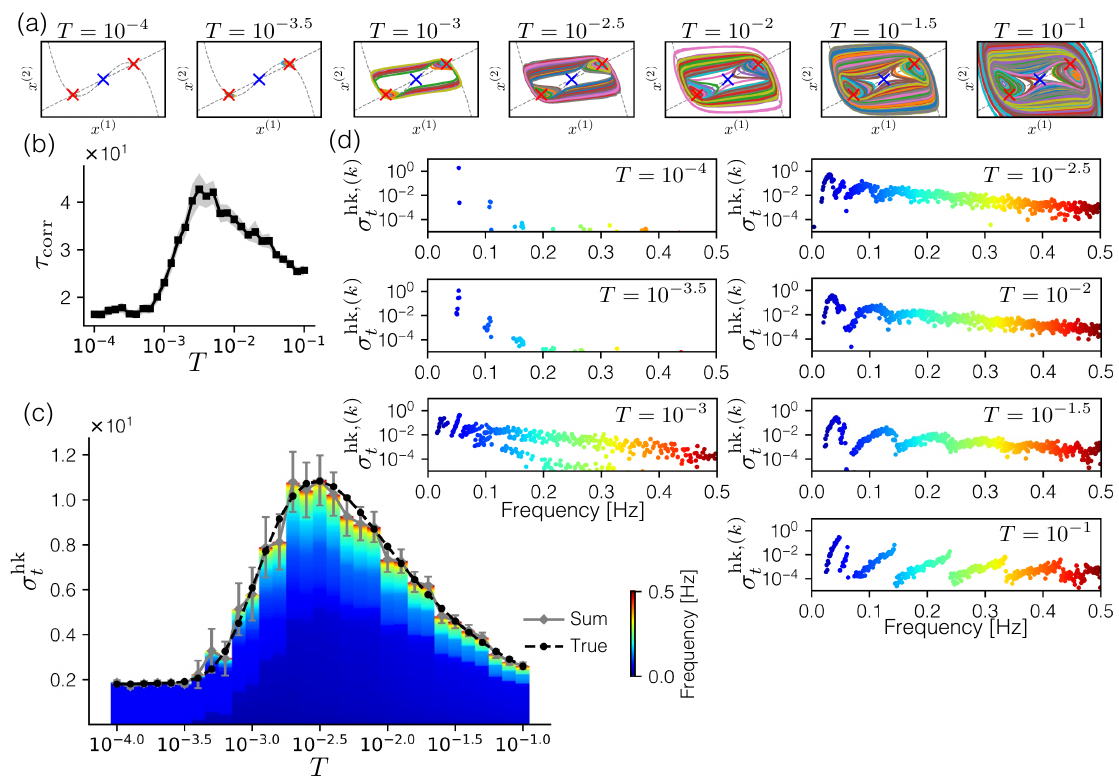}
    \caption{
    Our decomposition enables us to determine how the entropy production rate depends on the parameter $T$ in the context of the {\color{black} coherent resonance} in the noisy FitzHugh–Nagumo model.
    (a) Examples of the trajectory of the virtual dynamics in Eq.~\ref{eq:Langevin_hk} for the noisy FitzHugh--Nagumo model for different values of $T$. 
    The black dashed line represents the nullclines of the noisy FitzHugh--Nagumo model, which were calculated from the original Langevin dynamics in Eq.~\ref{eq:Langevin_FN} by ignoring the noise term. 
    The blue and red crosses represent the unstable and stable fixed points, respectively.
    For $T \leq 10^{-3.5}$, the trajectories are trapped in the small loop near the two fixed points. However, for $10^{-3.5}<T$, the trajectories transition between the two stable points, forming large loops in phase space. 
    (b) The correlation times $\tau_\textbf{corr}$, required to detect {\color{black} coherent resonance}. 
    The shaded areas represent 95\% confidence intervals. 
    (c) The parameter-dependent behavior of the housekeeping entropy production rate and its decomposition. 
    The stacked bar plot shows the sum of the contributions from the oscillatory modes. 
    The colors represent the frequencies of the oscillatory modes. 
    The gray line indicates the sum of the contributions from the oscillatory modes, with error bars representing 95\% confidence intervals. 
    The black dashed line shows the true housekeeping entropy production rate. 
    This curve exhibits an inverted-U shape, which is characteristic of {\color{black} coherent resonance}.  
    (d) The contribution of each oscillatory mode to the housekeeping entropy production rate $\sigma_t^{\mathrm{hk}, (k)}$ for different $T$. 
    When $T$ is small, only one frequency mode significantly contributes to the housekeeping entropy production rate $\sigma_t^\mathrm{hk}$. As $T$ increases, a broader range of frequency modes begins to contribute. 
    As the total entropy production rate begins to decrease at higher noise intensities, contributions from oscillatory modes gradually decrease.
    }
    \label{fig:FN_T}
\end{figure*}

Next, we applied our decomposition to investigate how oscillatory modes shape the housekeeping entropy production rate in {\color{black}a coherent-resonance regime of} the noisy FitzHugh--Nagumo model.
{\color{black}In this regime, noise induces temporally regular oscillatory behavior even in the absence of periodical external input \cite{pikovsky1997coherence, lindner2004effects}.}
{\color{black}This phenomenon is closely related to stochastic resonance as a form of noise-induced regularity.}

The virtual dynamics driven by the housekeeping local mean velocity $\bm{\nu}_t^\mathrm{hk}$ in Eq.~\ref{eq:Langevin_hk} of the noisy FitzHugh--Nagumo model with parameters $a=0$, $b=2$, $I=0$, and $\tau=12.5$ exhibit {\color{black}coherent resonance} as a function of the noise intensity $T$ (see Fig.~\ref{fig:FN_T}a).
With this parameter setting, the original Langevin system [Eq.~\ref{eq:Langevin_FN}] possesses two stable fixed points (see Fig.~\ref{fig:FN_T}a, red crosses).
When the noise intensity $T$ is small, the trajectories of the virtual dynamics remain confined near one of the stable points and rarely transition to the other.
As $T$ increases, the system begins to transition between the two stable points, forming large loops in phase space due to noise-induced switching. 
{\color{black}This noise-induced regularization of the dynamics} can also be quantitatively characterized by the correlation time {\color{black}$\tau_{\mathrm{corr}}$}, which measures the degree of temporal coherence in dynamical systems, as defined in Eq.~S54 in {\color{black} the Supporting Information, \textit{``Supplementary Methods''}}. 
As shown in Fig.~\ref{fig:FN_T}b, $\tau_{\mathrm{corr}}$ forms an inverted U-shaped curve with respect to the noise intensity $T$, indicating that both weak and strong noise lead to disordered dynamics, while an intermediate noise level produces the highest temporal coherence.

In line with the changes in trajectory structure described above, the total housekeeping entropy production rate, without applying our decomposition, also exhibits an inverted-U-shaped dependence on the noise intensity $T$, peaking around $T \simeq 10^{-2.5}$ (see Fig.~\ref{fig:FN_T}c).
It remains small when the system stays near one of the stable fixed points ($T\simeq 10^{-4}$), increases as transitions between the two basins emerge ($T\simeq 10^{-2.5}$), and decreases again under strong noise ($T\simeq 10^{-1}$), where the dynamics approach near-equilibrium behavior.
Although this overall trend reflects the characteristic signature of {\color{black} coherent resonance}, it does not reveal how individual oscillatory modes contribute to the underlying thermodynamic dissipation.

The underlying contributions of oscillatory modes to the housekeeping entropy production rate can again be revealed only through our decomposition; the observed peak in the housekeeping entropy production rate originates from broad frequency contributions (see Fig.~\ref{fig:FN_T}d).
At low noise intensities ($T\simeq 10^{-4}$), the entropy production rate is dominated by a single low-frequency mode corresponding to localized motion near a stable fixed point.
As the noise level increases ($T\simeq 10^{-2.5}$), additional frequency modes appear, showing that thermodynamic dissipation is distributed across a broader range of oscillatory components.
For even higher noise levels ($T\simeq 10^{-1}$), several frequency components selectively vanish, producing a harmonic-like, stepwise pattern in the modal spectrum.
These results highlight how our decomposition exposes frequency-resolved features of {\color{black} coherent resonance} that cannot be inferred from the total entropy production rate alone.

\section{Summary and discussion.}

This study improves our understanding of nonlinear oscillatory phenomena by providing a thermodynamic framework that quantifies the influence of oscillatory frequency and amplitude on the housekeeping entropy production rate (or the entropy production rate in the steady state) within noise-induced nonlinear systems.  Using the Koopman mode decomposition, we analyze the dissipation of oscillations in general nonlinear dynamics with noise by decomposing the housekeeping entropy production rate $\sigma_t^\mathrm{hk}$ into contributions from oscillatory modes. We applied this framework to the noisy FitzHugh–Nagumo model, using distinct parameter regimes corresponding to bifurcation and {\color{black} coherent resonance}, respectively. This revealed how Koopman modes of different frequencies contribute to the entropy production rate in the steady state of these nonlinear phenomena. Thus, the framework provides a clear explanation of dissipation in noise-induced nonlinear systems.

The present study provides a theoretical perspective for understanding how functional oscillations are organized and maintained under finite thermodynamic cost.
In living and synthetic systems, oscillatory dynamics are essential for realizing diverse functions ~\cite{van1928lxxii, noble1962modification, konopka1971clock, pittendrigh1960circadian, Buzsaki2006-se, Belousov1958-yu, zhabotinsky1964periodic} while operating within thermodynamic constraints~\cite{sekizawa2024decomposing, barato2017coherence, Del-Junco2020-px, Oberreiter2022-nx,ohga2023thermodynamic, shiraishi2023entropy, dechant2023thermodynamic, aguilera2025inferring, aslyamov2025excess,
Barato2016-lh, Cao2015-ho, Fei2018-ae, Lee2018-kb, nardini2017entropy, santolin2025dissipation, Uhl2019-ch, kolchinsky2024thermodynamic,zheng2024topological,  xu2025thermodynamic}.
The proposed framework reveals how individual oscillatory modes contribute to the overall thermodynamic dissipation, providing a structured basis for linking oscillatory dynamics with thermodynamic cost.
This perspective offers a means to examine how oscillatory systems achieve and regulate function under limited thermodynamic resources and may, in the future, help clarify how efficiency and design principles emerge across biological, physical, and engineered systems.

We clarify the scope and assumptions underlying our framework. 
In Eq.~\ref{eq:oscillation_x}, our framework is based on the diagonalizability of the Koopman generator under a finite-dimensional numerical approximation.
Although the Koopman generator is not generally diagonalizable~\cite{riechers2018beyond}, this treatment is justified when the number of Koopman modes is finite, because $\mathcal{K}$ is skew-adjoint in the virtual dynamics in Eq.~\ref{eq:Langevin_hk} (see Supporting Information, Section \textit{“Skew-adjointness and diagonalizability of the Koopman generator”}) and thus diagonalizable. 
However, this approach may not be applicable if there are infinitely many Koopman modes.
For example, chaotic systems are known to possess continuous spectra in their Koopman generators~\cite{arbabi2017ergodic, brunton2017chaos}, which may not be fully captured by a finite number of eigenvalues. 
Intuitively, this implies that chaotic systems cannot be expressed as a superposition of a finite number of oscillatory modes. 
It is unclear how valid Eq.~\ref{eq:oscillation_x} is when such a system is forced into a finite-dimensional approximation. 
If our framework is ineffective in systems with continuous spectra, this suggests that the virtual dynamics cannot be accurately represented by a finite set of oscillatory modes. 
Nevertheless, dynamic mode decomposition (DMD), which is a data-driven method to estimate the Koopman modes and eigenfunctions from time-series data, has been generalized to handle systems with continuous spectra~\cite{colbrook2023residual, lusch2018deep, brunton2017chaos, otto2019linearly}. Therefore, incorporating these approaches could enable us to extend our decomposition to systems for which a finite-dimensional approximation of the Koopman generator is not exactly valid.

{\color{black}
Although our analysis focuses on simplified dynamical models, Eq.~\ref{eq:bound} may provide insight into energetic constraints underlying neural oscillations. 
In the FitzHugh–Nagumo model, oscillatory dynamics are often interpreted as simplified representations of neuronal activity. 
The bound in Eq.~\ref{eq:bound} suggests that, if oscillatory modes correspond to neural rhythms, higher-frequency oscillations would require larger thermodynamic dissipation to achieve the same oscillation intensity. 
While the FitzHugh–Nagumo model is not intended to quantitatively describe metabolic energy consumption in neurons, this result highlights a potential thermodynamic constraint linking oscillation frequency, oscillation strength, and dissipative cost. 
Such constraints may offer a conceptual perspective on how energetic limitations could influence the observable frequencies and amplitudes of neural oscillations.

}

{\color{black}
A potential difficulty of the present approach is that estimation becomes more difficult in higher-dimensional chaotic systems.
One reason is that such systems can involve continuous or broadened spectral components, so a description based on only a small number of oscillatory modes may no longer be sufficient.
In addition, higher-dimensional chaotic systems can make it more difficult to simulate long trajectories of the virtual dynamics with sufficient numerical accuracy.
They can also make it harder to determine a state-space discretization that is fine enough to evaluate the underlying entropy production rate reliably.
For these reasons, the required number of modes and the numerical difficulty are expected to depend on the structure of the specific system, and we leave this issue for future work.

}

A promising direction for future research is to apply our decomposition framework to real data from noisy oscillatory systems. 
However, doing so would present practical challenges. 
Such an application would require addressing two issues simultaneously: (i) performing the Koopman mode decomposition of the housekeeping entropy production rate and (ii) carrying out the geometric decomposition of the velocity field into housekeeping and excess parts.
{\color{black}In this setting, finite-sample effects may further complicate both tasks, especially in low-noise metastable systems where rare transitions and low-probability but highly dissipative regions may be undersampled. Nevertheless, when this issue is not severe, both (i) and (ii) may still be tractable with existing methods.} 
Regarding (i), methods have been actively developed to identify Koopman eigenvalues, eigenfunctions and modes from time series data using data-driven approaches~\cite{williams2015data, lusch2018deep, otto2019linearly, takeishi2017learning, tu2014on, arbabi2017ergodic, brunton2017chaos, ichinaga2024pydmd, baddoo2023physics}. These methods may provide a foundation for implementing the Koopman analysis in practice.
Regarding (ii), recent advances have proposed methods for estimating the local mean velocity and its decomposition into housekeeping and excess parts from optimization problems based on thermodynamic uncertainty relations and optimal transport theory~\cite{li2019quantifying,otsubo2020estimating,otsubo2022estimating, dechant2022geometric, Dechant2022-gt, Ito2024-yt}. 
Thus, the housekeeping part of the velocity field may be obtained numerically from time series data via an optimization problem. Furthermore, incorporating the concept of inferring the lower bound on the entropy production rate under simplified assumptions regarding observables and interactions~\cite{lynn2022decomposing,lynn2022emergence, aguilera2025inferring} may also enable analysis of cases involving underlying complex dynamics.
While realizing both (i) and (ii) remains challenging, progress in both areas suggests that our decomposition could potentially be extended to analyze experimental data in the future.

\section{Materials and Methods}
\phantomsection
\makeatletter
\def\@currentlabelname{Materials and Methods}
\makeatother
\label{sec:Methods}
This section summarizes the numerical procedures used for the noisy FitzHugh–Nagumo model (Sec.~\textit{``\nameref{sec:fitzhugh}''}); see also the Supporting Information, \textit{``Supplementary Methods''}.
The numerical calculation consists of three steps:
(i) computing the housekeeping part of the local mean velocity $\bm{\nu}_t^\mathrm{hk}$ in the steady state,
(ii) simulating the virtual dynamics driven by $\bm{\nu}_t^\mathrm{hk}$ in Eq.~\ref{eq:Langevin_hk}, and
(iii) extracting Koopman eigenfunctions and modes from the generated time-series data and calculating the terms of our decomposition in Eq.~\ref{eq:decomposition_EP}.
The parameter settings were as follows. For Fig.~\ref{fig:FN_demo}a–e, we used $a=0$, $b=0.5$, $I=0$, $T=10^{-3}$, and $\tau=12.5$. For Fig.~\ref{fig:FN_demo}f–h, the same parameters were used except that $\tau$ was varied from $2.5$ to $22.5$ in increments of $5$. For Fig.~\ref{fig:FN_I}, we set $a=0$, $b=0.5$, $T=10^{-3}$, and $\tau=12.5$, while varying $I$ from $0$ to $2.5$ in increments of $0.1$. For Fig.~\ref{fig:FN_T}, we fixed $a=0$, $b=2$, $I=0$, and $\tau=12.5$, with $T$ varied from $10^{-4}$ to $10^{-1}$ in logarithmic steps of $0.1$.

\subsection{Calculation of the local mean velocity}
We calculated the housekeeping part of the local mean velocity $\bm{\nu}_t^\mathrm{hk}$ from the noisy FitzHugh--Nagumo model in Eq.~\ref{eq:Langevin_FN}.
The decomposition was performed in the steady state, where the excess part vanishes and $\bm{\nu}_t=\bm{\nu}_t^\mathrm{hk}$.
To compute $\bm{\nu}_t$, we estimated the steady-state distribution $p_t$, satisfying $\pdv{t} p_t(\bm{x})=0$, using the discretization approach of Ref.~\cite{gingrich2017inferring}: the continuous Langevin equation in Eq.~\ref{eq:Langevin_FN} was converted into a discrete transition-rate matrix on a grid.
We discretized $x^{(1)}$ and $x^{(2)}$ into $10^4$ intervals over $[-5,5]$ and $[-5+I,5+I]$, respectively, resulting in $10^{4\times 2}$ grid points.
In this discrete system, we constructed the transition-rate matrix corresponding to Eq.~\ref{eq:Langevin_FN} and obtained the steady-state distribution from the eigenvector associated with its zero eigenvalue.
From this distribution, we computed $\nabla \ln p_t$ by cubic-spline interpolation followed by differentiation, allowing us to evaluate the local mean velocity at arbitrary locations.

\subsection{Simulating the virtual dynamics}
Having obtained $\bm{\nu}_t^\mathrm{hk}$, we generated time-series trajectories by simulating the virtual dynamics in Eq.~\ref{eq:Langevin_hk}.
We used an eighth-order Runge–Kutta method with a time step $\Delta s = 1$, generating trajectories of length $S=150$.
For each parameter setting, we generated $N=1000$ independent trajectories to evaluate the decomposition.
These trajectories serve as Monte Carlo samples for approximating the terms in our decomposition, and their initial conditions were sampled from the discretized steady-state distribution $p_t(\bm{x})$.
Hereafter, $\bm{x}_{n,s}$ denotes the state at time $s$ of the $n$-th trajectory.

\subsection{Extraction of Koopman eigenfunctions and modes}
From the simulated time-series data, we estimated the Koopman eigenfunctions $\{\phi_{n, k}\}_{k=1}^{r_n}$, eigenvalues $\{\lambda_{n, k}\}_{k=1}^{r_n}$, and modes $\{\bm{v}_{n, k}\}_{k=1}^{r_n}$ for each trajectory.
Because different trajectories have different supports in state space, the eigenfunctions obtained from distinct trajectories were regarded as different functions.
Here, $r_n$ denotes the number of extracted modes for the $n$-th trajectory, and the double subscript $_{n,k}$ indicates the $k$-th eigenfunction, eigenvalue, or mode for that trajectory.
{\color{black}The number of extracted modes $r_n$ varies across trajectories, and the procedure for determining $r_n$ is described later.}
To obtain these quantities, we employed Hankel DMD~\cite{tu2014on, arbabi2017ergodic, brunton2017chaos} in combination with physics-informed DMD (PiDMD)~\cite{baddoo2023physics} via the PyDMD Python package~\cite{ichinaga2024pydmd}.
Hankel DMD constructs a vector of $h_n$ observable functions, $\bm{g}(\bm{x})=(\bm{g}_1(\bm{x})^\top, \bm{g}_2(\bm{x})^\top, \dots, \bm{g}_{h_n}(\bm{x})^\top)^\top=(\mathrm{Id}(\bm{x})^\top, (e^{\Delta s\mathcal{K}}\mathrm{Id}(\bm{x}))^\top, \dots, (e^{({h_n}-1)\Delta s\mathcal{K}}\mathrm{Id}(\bm{x}))^\top)^\top\in \mathbb{R}^{dh_n}$, where $\mathrm{Id}(\bm{x})=\bm{x}$ is the identity function.
Here, $h_n$ denotes the number of time delays chosen for the $n$-th trajectory, serving as a hyperparameter of the fitting procedure.
{\color{black}The procedure for selecting $h_n$ is also described later.}
This vector is obtained by time-delay embedding of the time-series data, i.e., $\bm{g}(\bm{x}_{n,s})=(\bm{x}_{n,s}^\top, \bm{x}_{n, s+\Delta s}^\top, \dots, \bm{x}_{n, s+({h_n}-1)\Delta s}^\top)^\top$.
Since the Koopman generator $\mathcal{K}$ is linear, the time evolution of this observable vector can be approximated by a linear dynamical system, even when the dynamics of $\bm{x}_{n,s}$ are nonlinear.

To estimate the Koopman generator $\mathcal{K}$ while ensuring that its eigenvalues are purely imaginary, we applied PiDMD, which constrains the representation matrix of $e^{\Delta s\mathcal{K}}$ to be unitary during the fitting procedure.
Let $L$ denote the representation matrix of $e^{\Delta s\mathcal{K}}$.
Under stationarity, the covariance matrix $\Sigma$ satisfies $\Sigma = L\Sigma L^*$, implying that $L$ is unitary when $\Sigma$ is the identity matrix.
Therefore, before applying PiDMD, we linearly transformed the delay-embedded data so that its covariance matrix became the identity matrix.
Specifically, we centered the delay-embedded observable vectors by subtracting their temporal mean, $\overline{\bm{g}} = (1/(S-h_n+1))\sum_{s=1}^{S-h_n+1} \bm{g}(\bm{x}_{n,s})$, and defined
$G = \qty(\bm{g}(\bm{x}_{n,1})-\overline{\bm{g}}, \ \bm{g}(\bm{x}_{n,2})-\overline{\bm{g}},\ \dots, \ \bm{g}(\bm{x}_{n,S-h_n+1})-\overline{\bm{g}})$.
We then performed a singular value decomposition $G = U S V^\top$, whitened the data as $\Tilde{G} = S^{-1} U^\top G$, and applied PiDMD to $\Tilde{G}$ to obtain the Koopman modes $\{\Tilde{\bm{v}}_{n,k}\}$.
{\color{black}Here, the number of extracted modes $r_n$ was determined for each trajectory from this singular value decomposition by applying optimal singular value hard thresholding (SVHT)~\cite{gavish2014optimal} to the singular values of $G$. 
Accordingly, $r_n$ varies across trajectories; for the example trajectory shown in Figs.~\ref{fig:FN_demo}b and c, we obtained $r_n=27$, whereas in analyses combining modes across trajectories, such as Fig.~\ref{fig:FN_demo}d, we used all extracted modes, giving a total of $r=\sum_{n=1}^N r_n$ modes. }Finally, the obtained modes were mapped back to the original coordinate system as $\bm{v}_{n,k} = U S \Tilde{\bm{v}}_{n,k}$, and the temporal mean $\overline{\bm{g}}$ was added back to the reconstructed trajectories so that they could be interpreted in the original observable space $\bm{g}(\bm{x}_{n,s})$.

For each trajectory, the number of time delays $h_n$ was chosen from 1 to 100 in increments of 1 to minimize the reconstruction error based on Eq.~\ref{eq:oscillation_x},
$\sum_{s=1}^S \left\|\bm{x}_{n,s} - \sum_{k=1}^{r_n} e^{\lambda_{n,k} s}\,\phi_{n,k}(\bm{x}_{n,0})\,\bm{v}_{n,k} \right\|^2$,
which varies with $h_n$ because the extracted eigenvalues $\lambda_{n,k}$, eigenfunctions $\phi_{n,k}$, and modes $\bm{v}_{n,k}$ depend on the chosen $h_n$.

\subsection{Computation of the terms of our decomposition}
From the Koopman eigenvalues, eigenfunctions, and modes obtained for each trajectory, we computed the terms of our decomposition as follows:
\begin{align}
    &\chi_{n,k} = |\lambda_{n,k}/(2\pi \mathrm{i})|\\
    &J_{n,k} = \frac{1}{S}\sum_s \qty(\phi_{n, k}\qty(\bm{x}_{n, s})\bm{v}_{n, k})^* D_t^{-1}\qty(\phi_{n, k}\qty(\bm{x}_{n, s})\bm{v}_{n, k}), \\
    &\sigma_t^{\mathrm{hk}, (n, k)} = (2\pi)^2\chi_{n, k}^2J_{n,k},
\end{align}
where the expected values in Eq.~\ref{eq:decomposition_EP} are approximated as time-averages within each trajectory.
Since each eigenfunction is supported only within its corresponding trajectory and does not overlap with those from other trajectories, the associated quantities $\chi_{n,k}$, $J_{n,k}$, and $\sigma_t^{\mathrm{hk}, (n,k)}$ were considered as distinct terms for each trajectory.
To summarize the results in a form consistent with ensemble expectations, we concatenated these quantities and reindexed them, dividing each trajectory-specific contribution by $N$:
$\{J_k\}_{k=1}^r = \{\, \frac{J_{1,1}}{N}, ..., \frac{J_{1,r_1}}{N}, ..., \frac{J_{N,1}}{N}, ..., \frac{J_{N,r_N}}{N} \,\}$,
$\{\sigma_t^{\mathrm{hk}, (k)}\}_{k=1}^r=\{\frac{\sigma_t^{\mathrm{hk}, (1, 1)}}{N}, ..., \frac{\sigma_t^{\mathrm{hk}, (1, r_1)}}{N}, ..., \frac{\sigma_t^{\mathrm{hk}, (N, 1)}}{N}, ..., \frac{\sigma_t^{\mathrm{hk}, (N, r_N)}}{N}\}$.
These reindexed quantities were plotted in Figs.~\ref{fig:FN_demo}-\ref{fig:FN_T}.
In frequency-resolved panels such as Fig.~\ref{fig:FN_demo}d, we used a moving window to reduce overlap and density bias across frequencies.
Values were summed within bins of width $10^{-3}$ and plotted at the bin centers.

\begin{acknowledgments}
The authors thank Shunsuke Kamiya and Isao Ishikawa for helpful discussions on Koopman analysis, and Ryuna Nagayama, Kohei Yoshimura, and Artemy Kolchinsky for fruitful discussions on geometric decomposition.
D.S. is supported by JSPS KAKENHI Grants No. 23KJ0799.
S.I. is supported by JSPS KAKENHI
Grants No. 22H01141, No. 23H00467, and No. 24H00834,
JST ERATO Grant No. JPMJER2302, and UTEC-UTokyo FSI Research Grant Program.
M.O. is supported by JST Moonshot R\&D Grant Number JPMJMS2012, and Japan Promotion Science, Grant-in-Aid for Transformative Research Areas Grant Numbers 23H04834. 
We used ChatGPT (OpenAI; GPT-5.4) and Gemini (Google; Gemini 3.1) for English proofreading and language editing during manuscript preparation. All scientific content was checked and approved by the authors.
\end{acknowledgments}

\bibliography{apssamp}

\clearpage
\onecolumngrid
\bigskip
\hrule
\begin{center}
    \vspace{1em}
    {\large \textbf{Supporting Information for}\\[3pt]
    \textit{Koopman Mode Decomposition of Thermodynamic Dissipation in Nonlinear Langevin Dynamics}}\\[6pt]
    Daiki Sekizawa, Sosuke Ito, and Masafumi Oizumi\\
    \vspace{1em}
\end{center}
\hrule
\bigskip


\DisableTOCfalse
\renewcommand{\thesection}{S\arabic{section}}
\renewcommand{\thesubsection}{(\Alph{subsection})}
\renewcommand{\theequation}{S\arabic{equation}}
\renewcommand{\thefigure}{S\arabic{figure}}
\setcounter{section}{0}\setcounter{equation}{0}\setcounter{figure}{0}

{

\tableofcontents
\bigskip
\hrule
\bigskip

\section{Koopman mode decomposition of virtual dynamics given by $\bm{\nu}_t^\mathrm{hk}$}
\phantomsection
\makeatletter
\def\@currentlabelname{Koopman mode decomposition of virtual dynamics given by $\bm{\nu}_t^\mathrm{hk}$} 
\makeatother
\label{sec:Koopman}

In this section, we provide a more detailed explanation of how the virtual deterministic process introduced in Eq.~\ref{eq:Langevin_hk} can be represented using Koopman mode decomposition. 
The virtual deterministic process is given by
\begin{align}
    d\bm{x}_s = \bm{\nu}_t^\mathrm{hk}(\bm{x}_s)\, ds.
    \label{eq_s:Langevin_hk_s}
\end{align}
As explained in the main text, the housekeeping part of the local mean velocity $\bm{\nu}_t^\mathrm{hk} (\bm{x})$ is derived from the original process [Eq.~\ref{eq:Langevin}] to obtain its housekeeping entropy production rate $\sigma_t^\mathrm{hk}$. 
In the virtual deterministic process in Eq.~\ref{eq_s:Langevin_hk_s}, the subscript $s$ denotes the virtual time, and $t$ denotes the time of the original Langevin dynamics. 
In this virtual dynamics, $\bm{\nu}_t^\mathrm{hk} (\bm{x})$ is fixed with respect to $s$, and the probability distribution $p_t(\bm{x})$ of the original dynamics serves as an invariant measure of the virtual dynamics.

The nonlinear dynamics in Eq.~\ref{eq_s:Langevin_hk_s} can be reformulated as a linear dynamical system in function space by extracting a finite number of modes via Koopman mode decomposition~\citeSI{koopman1931hamiltonian_, mezic2013analysis_} (see Fig.~\ref{fig:Koopman} in the main text). 
In continuous time, this is achieved using the Koopman generator $\mathcal{K}$, which is defined as the infinitesimal generator of the Koopman operator. 
For any observable $g: \mathbb{R}^d \to \mathbb{C}$, the generator acts as 
\begin{align}
    \mathcal{K} g(\bm{x}) := \nabla g(\bm{x}) \cdot  \bm{\nu}_t^{\mathrm{hk}}(\bm{x}).
    \label{def:koopman}
\end{align}
By definition, it satisfies
\begin{align}
    \mathcal{K} g(\bm{x}_s) &= \nabla g(\bm{x}_s) \cdot  \frac{d \bm{x}_s}{ds}= \frac{d}{ds} g(\bm{x}_s),
\end{align}
which describes the time evolution of the observable $g(\bm{x}_s)$. Moreover, $\mathcal{K}$ is linear. For any observables $g_1$ and $g_2$, and scalars $a$ and $b$, the relation
\begin{align}
    \mathcal{K}(a g_1 + b g_2) = a \mathcal{K} g_1 + b \mathcal{K} g_2,
\end{align}
holds. 
Thus, the Koopman generator converts the nonlinear dynamics of  Eq.~\ref{eq_s:Langevin_hk_s} into linear evolution in function space. For the identity observable $\mathrm{Id}(\bm{x})=\bm{x}$, its time evolution under Eq.~\ref{eq_s:Langevin_hk_s} can be written as
\begin{align}
    \bm{x}_{s+\Delta s} 
    = \mathrm{Id}(\bm{x}_{s+\Delta s}) 
    = e^{\Delta s\mathcal{K}}\, \mathrm{Id}(\bm{x}_{s}).
\end{align}
This equation demonstrates that the nonlinear dynamics of $\bm{x}$ are represented by a linear dynamical system in function space, with the identity observable $\mathrm{Id}(\cdot)$ as the initial condition.

Using the Koopman eigenfunctions as a basis for the function space enables us to understand that the nonlinear dynamics of $\bm{x}_s$ can be expressed as a sum of modal contributions in the virtual dynamics.  
We define the Koopman eigenfunctions $\{\phi_k\}_{k=1}^r$ and eigenvalues $\{\lambda_k\}_{k=1}^r$ as those that satisfy the following:
\begin{align}
    \mathcal{K} \phi_k(\bm{x}_s) = \lambda_k \phi_k(\bm{x}_s) = \frac{d}{ds}\phi_k(\bm{x}_s), 
\end{align}
where $r$ denotes the number of eigenvalues. The value of $r$ could be any integer, including infinity.
These eigenfunctions are solved as \begin{align}
    \phi_k(\bm{x}_{s+\Delta s}) = e^{\lambda_k\Delta s} \phi_k(\bm{x}_s).
\end{align}
Here, we assume that the number of modes $r$ is finite. Under this condition, as shown in Section \textit{``\nameref{sec:Skew-adjoint}''}, the Koopman generator $\mathcal{K}$ becomes diagonalizable.
Using this property, we expand the identity function $\mathrm{Id}(\bm{x})=\bm{x}$ with weight vectors $\{\bm{v}_k\}_{k=1}^r$ as
\begin{align}
    \mathrm{Id}(\bm{x}) = \sum_k^r \phi_k(\bm{x})\bm{v}_k. 
\end{align}
We then obtain the Koopman mode decomposition
\begin{align}
    \bm{x}_{s+\Delta s} &= \sum_k^r e^{\lambda_k \Delta s} \phi_k(\bm{x}_s) \bm{v}_k. \label{eq_s:oscillation_x}
\end{align}
The vector $\bm{v}_k$ is known as the Koopman mode. In this decomposition, the nonlinear dynamics of $\bm{x}_s$ is understood as a sum of the modal contributions. 
As the only time-dependent part is $e^{\lambda_k\Delta s}$, the Koopman eigenvalues $\{\lambda_k\}_{k=1}^r$ determine the characteristics of the modal dynamics. 
The real part of each eigenvalue determines the exponential growth or decay rate, while the imaginary part determines the oscillation frequency. 
In data analysis, dynamic mode decomposition offers efficient methods for extracting a finite number of modes from time series data. 

We define the frequency $\chi_k$ as \begin{align}
    \chi_k = |\lambda_k/(2\pi \mathrm{i})| \label{eq_s:def_chi},
\end{align} and $\mathrm{i}$ stands for the imaginary unit.  As will be derived in the next section, the eigenvalue $\lambda_k$ for Eq. \ref{eq_s:Langevin_hk_s} is purely imaginary, and therefore, $\chi_k$ is a real number. This implies that the time-variation of $\bm{x}_s$ in Eq. \ref{eq_s:oscillation_x} is expressed as the sum of oscillatory modes.

We also introduce the intensities of the oscillatory modes. When the eigenvalues are not degenerate, the intensity of the $k$-th oscillatory mode is given by \begin{align}
    J_k = \langle (\phi_k\bm{v}_k)^*D_t^{-1}(\phi_k\bm{v}_k)\rangle_t. \label{eq_s:def_normalized_intensity}
\end{align}
The symbol $^*$ stands for conjugate transpose. This quantity is the L2-norm of the $k$-th mode $\phi_k \bm{v}_k$ under the metric $D_t^{-1}p_t(\bm{x})$. Therefore, it represents the intensity of the $k$-th oscillatory mode.

\section{Skew-adjointness and diagonalizability of the Koopman generator}
\phantomsection
\makeatletter
\def\@currentlabelname{Skew-adjointness and diagonalizability of the Koopman generator}
\makeatother
\label{sec:Skew-adjoint}
To discuss the diagonalizability of the Koopman generator, we first examine its adjoint property in the context of the virtual dynamics driven by the housekeeping local mean velocity $\boldsymbol{\nu}_t^{\mathrm{hk}}(\boldsymbol{x})$. 
In this system, where $\nabla \cdot (\boldsymbol{\nu}^{\rm hk}_t(\boldsymbol{x}) p_t(\boldsymbol{x})) = 0$ holds, the Koopman generator $\mathcal{K}$ is a skew-adjoint operator, although this property does not hold for general dynamical systems.
For any functions $g_1(\boldsymbol{x})$ and $g_2(\boldsymbol{x})$, we can calculate $\langle g_2 \mathcal{K} g_1 \rangle_t= \int d\boldsymbol{x} g_2(\boldsymbol{x}) p_t(\boldsymbol{x}) \mathcal{K} g_1(\boldsymbol{x})$ as
\begin{align}
   \langle g_2 \mathcal{K} g_1 \rangle_t &= \int d\boldsymbol{x} g_2(\boldsymbol{x}) p_t(\boldsymbol{x}) \nabla g_1(\boldsymbol{x}) \cdot \boldsymbol{\nu}^{\rm hk}_t (\boldsymbol{x})  \nonumber \\
    &= -\int d\boldsymbol{x} g_1(\boldsymbol{x}) g_2(\boldsymbol{x}) \nabla  \cdot ( \boldsymbol{\nu}^{\rm hk}_t (\boldsymbol{x}) p_t(\boldsymbol{x}) ) - \int d\boldsymbol{x} g_1(\boldsymbol{x}) p_t(\boldsymbol{x}) \nabla  g_2(\boldsymbol{x}) \cdot  \boldsymbol{\nu}^{\rm hk}_t (\boldsymbol{x})  \nonumber \\
    &= -\langle g_1 \mathcal{K} g_2 \rangle_t,
\end{align}
where we used $\nabla  \cdot ( \boldsymbol{\nu}^{\rm hk}_t (\boldsymbol{x}) p_t(\boldsymbol{x}) )=0$ and applied integration by parts assuming that the distribution $p_t(\boldsymbol{x})$ becomes zero at infinity. 
Therefore, the relation $ \langle g_2 \mathcal{K} g_1 \rangle_t = - \langle g_1 \mathcal{K} g_2 \rangle_t$ shows that the Koopman generator is a skew-adjoint operator. This result implies that the matrix corresponding to $p_t(\boldsymbol{x})\mathcal{K}$ is antisymmetric when the Koopman generator is approximated numerically as a finite-dimensional matrix. Accordingly, the Koopman generator $\mathcal{K}$ is diagonalizable if the finite-dimensional approximation is sufficiently accurate.

\section{Koopman mode decomposition of the housekeeping entropy production rate}

\subsection{The main result}
As mentioned in the main text, our main result is a decomposition of the housekeeping entropy production rate into independent positive contributions from each oscillatory mode: 
\begin{align}
    \sigma_t^\mathrm{hk} &= \sum_k^r \sigma_t^{\mathrm{hk}, (k)} \nonumber\\
    \sigma_t^{\mathrm{hk}, (k)} &=(2\pi)^2\chi_k^2 J_k. \label{eq_s:decomposition_EP}
\end{align}
The decomposition means that the contribution of each oscillatory mode to the housekeeping entropy production rate is the product of its frequency squared $\chi_k^2$ and its intensity $J_k$ (see also Fig. \ref{fig:Koopman}b). In other words, modes with higher frequencies and greater intensities have a greater impact on the housekeeping entropy production rate.

\textcolor{black}{When the eigenvalues are degenerate, the decomposition becomes
\begin{align}
    \sigma_t^\mathrm{hk} &= \sum_{k'} (2\pi)^2\chi_{k'}^2 \left\langle \qty(\sum_{l\in C_{k'}}\phi_l\bm{v}_l)^*  {D_t^{-1}}\qty(\sum_{m\in C_{k'}}\phi_m\bm{v}_m)\right\rangle_t ,\label{eq_s:decomposition_EP_degenerate}
\end{align}
where $C_{k'}:=\{l \mid \lambda_l = 2\pi \chi_{k'} \mathrm{i}\}$ is the set of indices corresponding to the degenerate eigenvalue $2\pi i\chi_k$. Here, the index $k'$ is defined such that each $\chi_{k'}$ is distinct. The summation is taken over only those $k'$ for which $\chi_{k'}$ has different values, ensuring that no $k'$ with the same value of $\chi_{k'}$ is selected more than once.}

\subsection{Derivation of the main result}
\phantomsection
\makeatletter
\def\@currentlabelname{Derivation of the main result}
\makeatother
\label{sec:main_result_derivation}
We derive the main result [Eq.~\ref{eq_s:decomposition_EP}] using the Koopman mode decomposition, and the fact that the Koopman eigenvalues for the virtual dynamics in Eq.~\ref{eq_s:Langevin_hk_s} are purely imaginary. 

Using the Koopman mode decomposition [Eq.~\ref{eq_s:oscillation_x}], the housekeeping local mean velocity can be expressed as
\begin{equation}
    \bm{\nu}_t^\mathrm{hk}(\bm{x}_s)=\frac{d \bm{x}_s}{ds} = \sum_{k=1}^r \lambda_k \phi_k(\bm{x}_s) \bm{v}_k.
\end{equation}
The housekeeping entropy production rate $\sigma_t^\mathrm{hk}$ in the original dynamics [Eq.~\ref{eq:Langevin}] is then calculated as
\begin{align}
    \sigma^\mathrm{hk}_t
    &=\left\langle \qty(\bm{\nu}_t^\mathrm{hk})^* \mathit{D}_t^{-1}\bm{\nu}_t^\mathrm{hk} \right\rangle_t \nonumber\\
    &=\left\langle \qty(\sum_k \lambda_k \phi_k \bm{v}_k)^* \mathit{D}_t^{-1}\qty(\sum_l \lambda_l \phi_l \bm{v}_l) \right\rangle_t \nonumber\\
    &=\sum_{k, l} \lambda_k^* \lambda_l  \left\langle \qty(\phi_k \bm{v}_k)^* \mathit{D}_t^{-1}\qty(\phi_l \bm{v}_l)\right\rangle_t. \label{eq_s:derivation1}
\end{align}
Here, the symbol * is also used to represent the complex conjugate when applied to scalars.
First, we show that the Koopman eigenfunctions are orthogonal, i.e., $\langle \phi_k^*\phi_l\rangle_t=0$ if $\lambda_k\neq \lambda_l$. This orthogonality transforms Eq.~\ref{eq_s:derivation1} into our main result [Eq.~\ref{eq_s:decomposition_EP}]. To prove this orthogonality, we consider the following identity:
\begin{align}
    (\lambda_k^*+\lambda_l)\langle \phi_k^*\phi_l\rangle_t=&(\lambda_k^*+\lambda_l)\int d\bm{x}_s \ p_t(\bm{x}_s) \phi_k(\bm{x}_s)^*\phi_l(\bm{x}_s)\\
    =&\int d\bm{x}_s \ p_t(\bm{x}_s) \ \frac{d}{ds}(\phi_k(\bm{x}_s)^*\phi_l(\bm{x}_s))\nonumber\\
    =&\int d\bm{x}_s \ p_t(\bm{x}_s) \ \bm{\nu}_t^\mathrm{hk}\cdot \nabla(\phi_k(\bm{x}_s)^*\phi_l(\bm{x}_s))\nonumber\\
    =&-\int d\bm{x}_s \ [\nabla \cdot (p_t(\bm{x}_s) \bm{\nu}_t^\mathrm{hk}) ](\phi_k(\bm{x}_s)^*\phi_l(\bm{x}_s))\nonumber\\
    =&0, \label{eq_s:deribation2}
\end{align}
where we applied integration by parts and used the definition of the Koopman generator [Eq.~\ref{def:koopman}] and the definition of the housekeeping local mean velocity, i.e., $0 = -\nabla\cdot [\bm{\nu}_t^\mathrm{hk}(\bm{x})p_t(\bm{x})]$. 
From this identity [Eq.~\ref{eq_s:deribation2}], we obtain $\langle \phi_k^*\phi_l\rangle_t=0$ when $\lambda_k\neq \lambda_l$.

From this identity [Eq.~\ref{eq_s:deribation2}], we can also prove that the Koopman eigenvalues $\{\lambda_k\}_{k=1}^r$ are purely imaginary.
Substituting $k=l$ into Eq.~\ref{eq_s:deribation2} yields $\lambda_k^*+\lambda_k=0$, since $\langle|\phi_k|^2\rangle_t>0$. Therefore, all the eigenvalues are purely imaginary.

\section{Linear Langevin dynamics}
From our main result [Eq.~\ref{eq_s:decomposition_EP}], we can derive the decomposition for the linear Langevin process as a special case. This special case was obtained in our previous work~\citeSI{sekizawa2024decomposing_}. 

We consider the following linear Langevin process:
\begin{align}
    d\bm{x}_t = D_t A_t \bm{x}_t\, dt + \sqrt{2D_t}\, d\bm{B}_t.
    \label{eq_s:Langevin_linear}
\end{align}
Here, $A_t$ is a matrix representing the linear dynamics.  We assume that the distribution $p_t(\bm{x})$ is Gaussian. In order to  apply our decomposition, we consider the virtual dynamics
\begin{align}
    d\bm{x}_s
    = \bm{\nu}_t^\mathrm{hk}(\bm{x}_s)\, ds
    = D_t A_t^\mathrm{hk}\, \bm{x}_s\, ds,
\end{align}
where $A_t^\mathrm{hk}$ is a matrix discussed in Ref.~\citeSI{sekizawa2024decomposing_}. The matrix $A_t^\mathrm{hk}$ exists if the original Langevin process in Eq.~\ref{eq_s:Langevin_linear} is linear and  if the distribution $p_t(\bm{x})$ is Gaussian. We note that the real matrix $D_t A_t^{\rm hk}$ can be expressed as the product of a real antisymmetric matrix and a positive-definite symmetric matrix (see Ref.~\citeSI{sekizawa2024decomposing_}), and is diagonalizable. Because $D_t A_t^{\rm hk}$ is diagonalizable, this virtual dynamics is solved as follows:
\begin{align}
    \bm{x}_{s+\Delta s}
    &= e^{D_t A_t^\mathrm{hk}\Delta s}\, \bm{x}_s \\
    &= \sum_k e^{\lambda_k \Delta s}\, \mathsf{P}\bm{e}_k \bm{e}_k^\top \mathsf{P}^{-1}\bm{x}_s \\
    &= \sum_k e^{\lambda_k \Delta s}\, \mathsf{F}_k\, \bm{x}_s,
    \label{eq_s:oscillation_linear}
\end{align}
where $\lambda_k$ is the $k$th eigenvalue of $D_t A_t^\mathrm{hk}$ and we consider the eigendecomposition of $D_t A_t^\mathrm{hk}$:
\begin{align}
    D_t A_t^\mathrm{hk}
    &= \mathsf{P}\mathsf{\Lambda}\mathsf{P}^{-1}
     = \sum_k \lambda_k\, \mathsf{P}\bm{e}_k \bm{e}_k^\top \mathsf{P}^{-1}
     = \sum_k \lambda_k\, \mathsf{F}_k, \\
    \mathsf{F}_k
    &= \mathsf{P}\bm{e}_k \bm{e}_k^\top \mathsf{P}^{-1}.
\end{align}
The matrix $\mathsf{P}$ is regular and complex-valued. The matrix $\mathsf{F}_k$ is regarded as the projection matrix. The matrix $\mathsf{\Lambda}$ is a diagonal matrix with $k$-th \emph{entry} being the $k$-th eigenvalue $\lambda_k$.  The vector $\bm{e}_k$  has a value of $1$ in the $k$-th position and $0$ in all other positions.

We can relate this solution to the expression in Eq.~\ref{eq_s:oscillation_x}, which uses Koopman eigenfunctions and modes, by considering the following quantities:
\begin{align}
    \phi_k(\bm{x}) &= \bm{e}_k^\top \mathsf{P}^{-1}\bm{x}, \\
    \bm{v}_k       &= \mathsf{P}\bm{e}_k.
\end{align}
Because the Koopman generator $\mathcal{K}$ is given by $\mathcal{K} g(\boldsymbol{x})=\nabla g(\boldsymbol{x}) \cdot \boldsymbol{\nu}^{\rm hk}_t(\boldsymbol{x})=\nabla g(\boldsymbol{x}) \cdot D_t A_t^\mathrm{hk}\, \bm{x}$, we obtain $\mathcal{K} \phi_k(\boldsymbol{x})=( (\mathsf{P}^{-1})^{\top}\bm{e}_k )  \cdot \mathsf{P}\mathsf{\Lambda}\mathsf{P}^{-1} \, \bm{x} = \lambda_k \bm{e}_k^\top \mathsf{P}^{-1}\bm{x} = \lambda_k \phi_k(\boldsymbol{x}) $. Therefore, the $k$-th eigenvalue $\lambda_k$ of the matrix $D_t A_t^\mathrm{hk}$ is regarded as the $k$-th Koopman eigenvalue.  By substituting the eigenvalue $\lambda_k$, eigenfunction $\phi_k$, and Koopman mode $\bm{v}_k$ into our decomposition in Eq.~\ref{eq_s:decomposition_EP}, we obtain the result presented in Ref.~\citeSI{sekizawa2024decomposing_} as a special case:
\begin{align}
    \sigma_t^\mathrm{hk}
    &= \sum_k |\lambda_k|^2 \bigl\langle (\phi_k \bm{v}_k)^* D_t^{-1} (\phi_k \bm{v}_k)\bigr\rangle_t \nonumber\\
    &= \sum_k |\lambda_k|^2 \bigl\langle ((\bm{e}_k^\top \mathsf{P}^{-1}\bm{x}_t)(\mathsf{P}\bm{e}_k))^* D_t^{-1} ((\bm{e}_k^\top \mathsf{P}^{-1}\bm{x}_t)(\mathsf{P}\bm{e}_k))\bigr\rangle_t \notag \\
    &= \sum_k |\lambda_k|^2 \bigl\langle (\mathsf{F}_k \bm{x}_t)^* D_t^{-1} (\mathsf{F}_k \bm{x}_t)\bigr\rangle_t.
\end{align}

{\color{black}

\section{Physical meaning of frequencies extracted in our methods}

In this section, we further examine the oscillatory frequencies extracted by our decomposition in order to clarify their physical meaning.
Specifically, we compare the frequencies extracted by our decomposition with three established notions of oscillation in dynamical systems:
(A) the characteristic frequency predicted by linear stability analysis around a stable fixed point,
(B) the oscillation frequency of a deterministic limit cycle, and
(C) the frequencies associated with the spectrum of the corresponding stochastic (Fokker--Planck) operator.

A common point underlying all three comparisons is that the frequencies extracted by our decomposition are determined not by the deterministic drift $D_t\bm{f}_t(\bm{x})$ alone, which governs the zero-noise dynamics, but by the local mean velocity field
\begin{align}
\bm{\nu}_t(\bm{x}) = D_t\bm{f}_t(\bm{x}) - D_t\nabla \ln p_t(\bm{x}),
\end{align}
which includes both the deterministic drift and the diffusion-induced probability transport.
Accordingly, as long as the noise is nonzero, even if it is arbitrarily small, the extracted frequencies should be interpreted as frequencies of the stochastic probability flow described by $\bm{\nu}_t$.
This also implies that their relation to the oscillatory structure of the zero-noise deterministic dynamics is not universal, but depends on the situation.

For the comparisons in (A) and (B), we use the noisy FitzHugh--Nagumo model with parameter regimes chosen based on the eigenvalues of the Jacobian of the deterministic drift $D_t\bm{f}_t$.
Figure~\ref{fig:linear_stability_analsis}a shows how these eigenvalues change as the parameter $I$ varies.
Based on this analysis, we select representative parameter regimes in which the deterministic dynamics either exhibit a limit cycle or admit a stable fixed point, as illustrated in Fig.~\ref{fig:linear_stability_analsis}.
For (C), we consider the linear Langevin setting, where the relationship to the stochastic generator spectrum can be stated explicitly.
The analysis methods used in this section are described in Section \textit{``\nameref{sec:Methods_for_SI}''}.

\begin{figure}
    \centering
    \includegraphics[width=\linewidth]{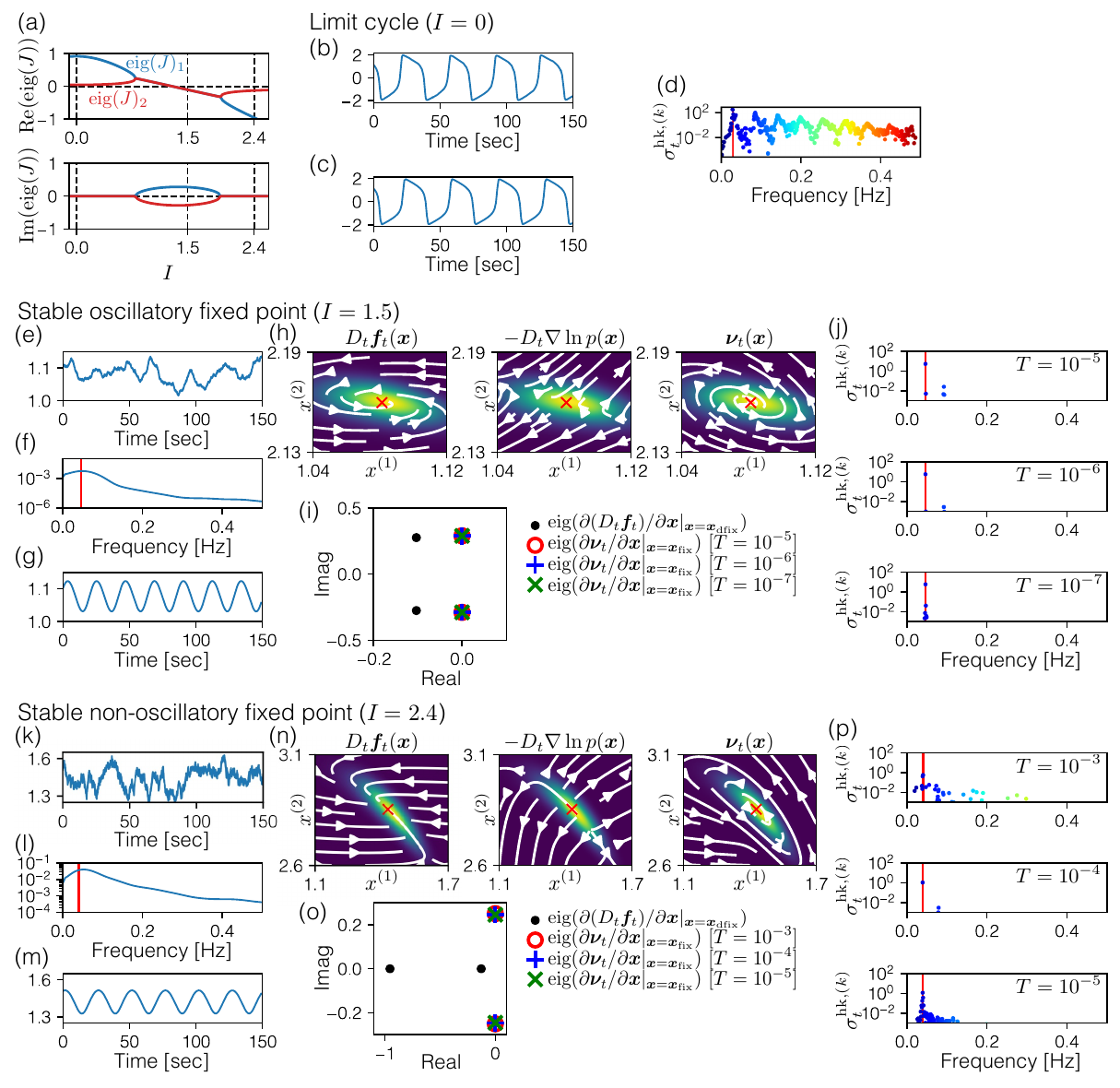}
    \caption{{\color{black}
    Comparison between frequencies extracted by our decomposition and those characterized by limit-cycle dynamics and linear stability analysis.
    (a) Eigenvalues of the Jacobian of the deterministic drift $D_t\bm{f}_t(\bm{x})$ evaluated at the fixed point $\boldsymbol{x}_{\rm fix}$ as a function of the input current $I$, used to select representative parameter regimes: a limit-cycle regime ($I=0$), a stable oscillatory fixed point ($I=1.5$), and a stable non-oscillatory fixed point ($I=2.4$).
    (b,c) Time series of $x_t^{(1)}$ obtained from the Langevin dynamics (b) and the corresponding virtual dynamics driven by $\bm{\nu}_t^\text{hk}$ (c) in the limit-cycle regime ($I=0$).
    (d) Frequency-resolved contributions to the entropy production rate obtained by our decomposition for $I=0$; the red vertical line indicates the limit cycle frequency $f_{\text{LC}}$, which is defined in Section \textit{``\nameref{sec:Methods_for_SI}''}.
    (e-j) Stable oscillatory fixed-point regime ($I=1.5$).
    (e) Time series of $x_t^{(1)}$ from the Langevin dynamics.
    (f) Power spectral density of the time series in (e), estimated using Welch's method; the red vertical line indicates the frequency given by the imaginary part of the Jacobian eigenvalues of $\bm{\nu}_t^\text{hk}$.
    (g) Time series generated by the virtual dynamics.
    (h) Vector fields of the deterministic drift $D_t\bm{f}_t(\bm{x})$ (left), the diffusion-induced term $-D_t\nabla\ln p_t(\bm{x})$ (middle), and the local mean velocity field $\bm{\nu}_t^\text{hk}$ (right); color indicates the steady-state probability density $p_t(\bm{x})$, and the red cross marks the deterministic fixed point of $D_t\bm{f}_t$.
    (i) Eigenvalues of the Jacobian of $D_t\bm{f}_t(\bm{x})$ at the deterministic fixed point $\boldsymbol{x}_{\rm dfix}$ and the Jacobian of $\bm{\nu}_t^\text{hk}$ at the fixed point $\boldsymbol{x}_{\rm fix}$ for different noise intensities.
    (j) Frequency-resolved contributions to the entropy production rate. 
    The red vertical line represents the frequency $f_{\text{LSA}, \bm{\nu}_t}$, which is determined from the linear stability analysis of $\bm{\nu}_t^\text{hk}$ and defined in Section \textit{``\nameref{sec:Methods_for_SI}''}.
    (k-p) Stable non-oscillatory fixed-point regime ($I=2.4$), shown in the same format as in (e-j).
    }}
    \label{fig:linear_stability_analsis}
\end{figure}

The results can be summarized as follows.
Around a stable fixed point (A), the characteristic frequency relevant to our decomposition is determined not by the Jacobian of the deterministic drift $D_t\bm{f}_t$ alone, but by that of the local mean velocity field $\bm{\nu}_t^{\mathrm{hk}}$, which incorporates the effect of stochastic diffusion through the probability distribution.
Accordingly, the frequencies predicted from the Jacobian of $\bm{\nu}_t^{\mathrm{hk}}$ agree with those extracted by our decomposition.
In the limit-cycle regime (B), the dominant frequency extracted by our decomposition coincides with the oscillation frequency of the stochastic dynamics in the representative small-noise regime considered here.
Based on these observations, we then discuss in (C) how the frequencies extracted by our decomposition relate to those associated with the stochastic (Fokker--Planck) operator.
Since the spectral frequencies of the Fokker--Planck operator are determined by the deterministic drift $D_t\bm{f}_t$, they do not necessarily coincide with those extracted by our decomposition.

\subsection{Comparison with frequencies extracted by linear stability analysis}

We next examine parameter regimes in which the deterministic drift admits a stable fixed point.
In such regimes, and for sufficiently small noise, the dynamics are restricted to fluctuations around the fixed point.
As shown in Fig.~\ref{fig:linear_stability_analsis}a, we consider two representative cases: one in which the Jacobian
\begin{align}
J := \left. \frac{\partial (D_t \bm{f}_t)}{\partial \bm{x}} \right|_{\boldsymbol{x}=\boldsymbol{x}_{\rm dfix}}
\end{align}
evaluated at the deterministic fixed point $\boldsymbol{x}_{\rm dfix}$ has a complex conjugate pair of eigenvalues with negative real parts ($I = 1.5$), and another in which all eigenvalues are purely real and negative ($I = 2.4$). Here, the deterministic fixed point $\boldsymbol{x}_{\rm dfix}$ is defined as $D_t \bm{f}_t(\boldsymbol{x}_{\rm dfix}) =\bm{0}$.

The results show that, in these fixed-point regimes, the frequencies extracted by our decomposition are consistently explained by the linear stability analysis of the local mean velocity field $\bm{\nu}_t^{\mathrm{hk}}$, whereas agreement with the corresponding analysis of the deterministic drift occurs only in some cases.
To make this point explicit, we compare the two representative fixed-point regimes introduced above.
In (A-1), the Jacobian of the deterministic drift already has complex eigenvalues, so the deterministic dynamics contain a damped oscillatory tendency.
In this case, the imaginary part of the eigenvalues of the Jacobian of $D_t\bm{f}_t(\bm{x})$ happens to agree with the dominant frequency extracted by our decomposition, and the frequency predicted from the Jacobian of the local mean velocity field $\bm{\nu}_t^{\mathrm{hk}}$ is also consistent with our decomposition.
In (A-2), by contrast, the Jacobian of $D_t\bm{f}_t(\bm{x})$ has only real eigenvalues, so at the level of the deterministic drift there is no oscillatory frequency to compare with.
Nevertheless, in both regimes, the Jacobian of the local mean velocity field $\bm{\nu}_t^{\mathrm{hk}}$ has imaginary eigenvalues, and the corresponding predicted frequencies are consistent with those extracted by our decomposition.
Thus, in this noisy FitzHugh--Nagumo example, our decomposition does not characterize oscillations of the deterministic drift alone, but rather oscillations of the effective probability flow that includes diffusion-induced effects.

\subsubsection*{(A-1) Stable oscillatory fixed point}

We first consider the case $I=1.5$, and show that even though the deterministic drift $D_t\bm{f}_t(\bm{x})$ around the stable fixed point has only a damped oscillatory tendency, the stochastic dynamics exhibit an oscillation that remains sustained around the fixed point by diffusion-induced effects.
In this regime, the deterministic drift $D_t\bm{f}_t(\bm{x})$ has a stable fixed point with a complex-conjugate pair of Jacobian eigenvalues.
Consistently, the Langevin trajectory exhibits stochastic oscillations (Fig.~\ref{fig:linear_stability_analsis}e), the power spectral density shows a clear peak (Fig.~\ref{fig:linear_stability_analsis}f), and the virtual dynamics generated by $\bm{\nu}_t^{\mathrm{hk}}$ also exhibit sustained oscillations with the same frequency (Fig.~\ref{fig:linear_stability_analsis}g).

The velocity fields clarify why such oscillatory motion persists in the stochastic dynamics even when the noise is arbitrarily small but nonzero.
As shown in Fig.~\ref{fig:linear_stability_analsis}h, the deterministic drift $D_t\bm{f}_t(\bm{x})$ alone drives trajectories toward the stable fixed point and therefore does not by itself sustain oscillation, whereas the diffusion-induced term $-D_t\nabla\ln p_t(\bm{x})$ modifies the effective velocity field so that the resulting local mean velocity field $\bm{\nu}_t^{\mathrm{hk}}$ exhibits a circulating probability flow around the fixed point.
Figure~\ref{fig:linear_stability_analsis}i shows that the Jacobian of $\bm{\nu}_t^{\mathrm{hk}}$ retains an oscillatory frequency even though the deterministic drift still drives trajectories toward the stable fixed point.

The frequency extracted by our decomposition should therefore be compared with the Jacobian of $\bm{\nu}_t^{\mathrm{hk}}$ rather than with that of the deterministic drift alone.
As shown in Fig.~\ref{fig:linear_stability_analsis}j, the oscillation frequency predicted from the imaginary part of the eigenvalues of the Jacobian of $\bm{\nu}_t^{\mathrm{hk}}$ agrees with the dominant frequency extracted by our decomposition.
In this regime, the Jacobian of $\bm{\nu}_t^{\mathrm{hk}}$ is close to that of the deterministic drift, which is why the frequency predicted from $D_t\bm{f}_t(\bm{x})$ also happens to take a similar value.
Thus, even in this case, the relevant oscillation extracted by our decomposition is most naturally interpreted as that of the effective probability flow described by $\bm{\nu}_t^{\mathrm{hk}}$, while the agreement with the deterministic drift is only incidental.
This distinction becomes fully explicit in (A-2), where the deterministic drift no longer provides any oscillatory prediction.

\subsubsection*{(A-2) Stable non-oscillatory fixed point}

We next consider the case $I=2.4$, and show more sharply that even though the deterministic drift $D_t\bm{f}_t(\bm{x})$ around the stable fixed point has no oscillatory tendency at all, the oscillation extracted by our decomposition remains sustained around the fixed point by diffusion-induced effects.
In this regime, the Jacobian of the deterministic drift $D_t\bm{f}_t(\bm{x})$ has only real negative eigenvalues, so at the level of the deterministic drift there is no oscillatory frequency to compare with.
Nevertheless, the Langevin trajectory exhibits stochastic oscillations (Fig.~\ref{fig:linear_stability_analsis}k), the power spectral density shows a clear peak (Fig.~\ref{fig:linear_stability_analsis}l), and the virtual dynamics generated by $\bm{\nu}_t^{\mathrm{hk}}$ also exhibit sustained oscillations with the same frequency (Fig.~\ref{fig:linear_stability_analsis}m).

The velocity fields clarify why such oscillatory motion persists in the stochastic dynamics.
As shown in Fig.~\ref{fig:linear_stability_analsis}n, the deterministic drift $D_t\bm{f}_t(\bm{x})$ still drives trajectories toward the stable fixed point, but the relaxation occurs along curved paths in phase space because different directions relax at different rates.
When stochastic diffusion is present, random perturbations repeatedly displace the state across these curved deterministic flows, and the diffusion-induced term $-D_t\nabla\ln p_t(\bm{x})$ modifies the effective velocity field so that the resulting local mean velocity field $\bm{\nu}_t^{\mathrm{hk}}$ exhibits a circulating probability flow around the fixed point.
Figure~\ref{fig:linear_stability_analsis}o shows that the Jacobian of $\bm{\nu}_t^{\mathrm{hk}}$ acquires an oscillatory frequency even though the Jacobian of the deterministic drift has only real eigenvalues.

This distinction persists even in the infinitesimal-noise limit because the diffusion-induced contribution does not disappear from the Jacobian of the effective probability flow.
Around the fixed point $\bm{x}_{\mathrm{fix}}$, the local mean velocity field is approximated by
\begin{align}
\bm{\nu}_t(\bm{x})
\simeq
(J + D_t\Sigma^{-1})(\bm{x}-\bm{x}_{\mathrm{fix}}),
\end{align}
where $J$ is the Jacobian of the deterministic drift and $\Sigma$ is the covariance matrix of the stationary distribution near the fixed point.
For linear Langevin dynamics, $\Sigma$ satisfies
\begin{align}
J\Sigma + \Sigma J^\top + 2D_t = 0 .
\end{align}
Since $\Sigma$ scales proportionally to the noise intensity, the product $D_t\Sigma^{-1}$ remains finite even when the noise intensity becomes arbitrarily small.
Therefore, even though the Jacobian of the deterministic drift has only real eigenvalues, the Jacobian of $\bm{\nu}_t$ can acquire imaginary eigenvalues.

The frequency extracted by our decomposition should therefore be compared with the Jacobian of $\bm{\nu}_t^{\mathrm{hk}}$ rather than with that of the deterministic drift alone.
Because the Jacobian of $D_t\bm{f}_t(\bm{x})$ has no imaginary part, the deterministic drift cannot explain the nonzero frequency extracted by our decomposition in this regime.
By contrast, Fig.~\ref{fig:linear_stability_analsis}p shows that the oscillation frequency predicted from the imaginary part of the eigenvalues of the Jacobian of $\bm{\nu}_t^{\mathrm{hk}}$ agrees with the dominant frequency extracted by our decomposition.
Thus, even more clearly than in (A-1), the relevant oscillation extracted by our decomposition is most naturally interpreted as that of the effective probability flow described by $\bm{\nu}_t^{\mathrm{hk}}$, while the deterministic drift provides no corresponding oscillatory prediction.

\subsection{Comparison with the limit-cycle frequency in a small-noise regime}

We next consider how the frequencies extracted by our decomposition compare with the actual oscillation frequency in the limit-cycle regime.
As shown in Fig.~\ref{fig:linear_stability_analsis}a, we select the representative limit-cycle regime at $I=0$.
In this regime, simulations of the Langevin dynamics exhibit clear nonlinear oscillations (Fig.~\ref{fig:linear_stability_analsis}b).
Because the noise intensity is small, the corresponding virtual dynamics driven by the local mean velocity field $\bm{\nu}_t^\text{hk}$ also display oscillatory behavior with a similar characteristic frequency (Fig.~\ref{fig:linear_stability_analsis}c).

We then compare the oscillation frequency extracted from the Langevin time series with the frequency-resolved decomposition of the entropy production rate.
As shown in Fig.~\ref{fig:linear_stability_analsis}d, the dominant contribution to the entropy production rate appears at the same frequency as the oscillation frequency of the Langevin limit cycle.
Thus, in the representative small-noise limit-cycle regime considered here, the dominant frequency extracted by our decomposition agrees with the actual oscillation frequency of the stochastic dynamics.

\subsection{Comparison with frequencies extracted by the stochastic operator}

We finally discuss how the frequencies extracted by our decomposition relate to those associated with the stochastic (Fokker--Planck) operator.
Based on the fixed-point analyses in Section~(A), this comparison can be stated explicitly in the linear Langevin setting.
In this setting, the frequencies associated with the Fokker--Planck operator are determined by the deterministic drift $D_t\bm{f}_t$, whereas the frequencies extracted by our decomposition are determined by the local mean velocity field $\bm{\nu}_t^{\mathrm{hk}}$.
Accordingly, as already seen in Section~(A), the two can coincide in some situations, but they need not do so in general, because diffusion-induced effects can modify the effective probability flow and thereby shift the frequencies extracted by our decomposition away from those associated with the deterministic drift.
Beyond this linear setting, we do not currently know a universal relation in general nonlinear systems.

The eigenvalues of the Fokker--Planck operator are often used to characterize oscillatory behavior in stochastic systems because the time evolution of the probability density can be expanded in terms of its eigenfunctions.
To examine this relation, consider the linearized Langevin dynamics around a stable fixed point,
\begin{align}
d\bm{x}
=
J(\bm{x}-\bm{x}_\text{fix})\,dt
+
\sqrt{2D_t}\,d\bm{B}_t .
\end{align}
If $\mu_1,\dots,\mu_d$ denote the eigenvalues of the Jacobian $J$ in a $d$-dimensional system, then the spectrum of the corresponding Fokker--Planck operator $\mathcal{L}_{\mathrm{FP}}$ is given by finite sums of the eigenvalues of $J$
as discussed in Section \textit{``\nameref{sec:FP_eigvals}''}.
When the fixed point is stable so that $\mathrm{Re}(\mu_i)\le 0$ with equality only for the stationary mode, the eigenvalues whose real parts are closest to zero correspond directly to the eigenvalues $\mu_i$ themselves.
The imaginary parts of these eigenvalues therefore determine the frequency of the most persistent oscillatory component of the stochastic dynamics.

By contrast, the frequencies extracted by our decomposition reflect oscillatory modes of the effective probability flow including the diffusion-induced contribution.
As discussed in Section~(A), these oscillatory modes are associated with the Jacobian of the local mean velocity field $\bm{\nu}_t^{\mathrm{hk}}$, which can differ from that of the deterministic drift.
Therefore, even within the linear regime, the relation between the frequencies extracted by our decomposition and those associated with the stochastic generator is not one of universal equality, but is controlled by the difference between $\bm{\nu}_t^{\mathrm{hk}}$ and $D_t\bm{f}_t$.
When the Jacobians of $\bm{\nu}_t^{\mathrm{hk}}$ and $D_t\bm{f}_t$ are close, the two frequencies are similar.
When diffusion-induced effects substantially modify the effective velocity field, they differ.

Taken together, these results provide the following physical interpretation of the frequencies extracted by our decomposition.
In the limit-cycle regime considered here, the dominant frequency extracted by our decomposition agrees with the actual oscillation frequency of the stochastic dynamics.
In the fixed-point regimes considered here, the extracted frequencies represent oscillatory motions of the stochastic probability flow sustained by diffusion-induced effects around the stable fixed point, and are consistently predicted by the linear stability analysis of $\bm{\nu}_t^{\mathrm{hk}}$.
In the linear setting, the comparison with the Fokker--Planck operator further shows that these frequencies are tied to the effective probability flow rather than, in general, to the deterministic drift alone.

\section{Eigenvalues of the Fokker--Planck operator for linear Langevin dynamics}
\phantomsection
\makeatletter
\def\@currentlabelname{Eigenvalues of the Fokker--Planck operator for linear Langevin dynamics}
\makeatother
\label{sec:FP_eigvals}

In this section, to help clarify the relation between the stochastic operator and our decomposition in Section \textit{``Physical meaning of frequencies extracted in our methods,''} we provide an intuitive demonstration of the association between the eigenvalues of the Fokker–Planck operator and linear Langevin dynamics. For a detailed proof and explanations, see Refs.~\citeSI{liberzon2000spectral_, leen2016eigenfunctions_}.

Consider
\begin{align}
d\bm{x}_t = A\bm{x}_t\,dt + \sqrt{2D}\,d\bm{B}_t ,
\end{align}
where \(A\in\mathbb{R}^{d\times d}\), \(D\in\mathbb{R}^{d\times d}\) is symmetric positive semidefinite, and \(\bm{B}_t\) is a \(d\)-dimensional standard Brownian motion.
The corresponding Fokker--Planck operator $\mathcal L_{\mathrm{FP}}$ is defined as
\begin{align}
\mathcal L_{\mathrm{FP}} p_t
=
-\sum_i \partial_{x_i}\!\left(\sum_j A_{ij}x_j\,p_t\right)
+
\sum_{i,j}\partial_{x_i}\!\left(D_{ij}\partial_{x_j}p_t\right).
\end{align}
If all eigenvalues of \(A\) have negative real parts, then the process admits a stationary Gaussian density
\begin{align}
\rho_{\mathrm{ss}}(\bm{x})
=
\frac{1}{Z}
\exp\!\left(
-\frac12 \bm{x}^\top\Sigma^{-1}\bm{x}
\right),
\end{align}
where \(\Sigma\) satisfies
\begin{align}
A\Sigma+\Sigma A^\top+2D=0 .
\end{align}

The key simplification is that \(\mathcal L_{\mathrm{FP}}\) preserves the form \(P\rho_{\mathrm{ss}}\), where \(P\) is a polynomial in $\boldsymbol{x}$.
Indeed, using \(\nabla\rho_{\mathrm{ss}}=-\Sigma^{-1}\bm{x}\rho_{\mathrm{ss}}\) and \(\mathcal L_{\mathrm{FP}}\rho_{\mathrm{ss}}=0\), one finds
\begin{align}
\mathcal L_{\mathrm{FP}}(P\rho_{\mathrm{ss}})
=
\rho_{\mathrm{ss}}
\left(
(\tilde{A}\bm{x})\cdot\nabla P
+
\sum_{i,j}D_{ij}\partial_{x_i}\partial_{x_j}P
\right),
\label{eq_s:FP_conjugated_short}
\end{align}
where
\begin{align}
\tilde{A}:=\Sigma A^\top \Sigma^{-1}.
\end{align}
Here, \(\tilde{A}\) has the same eigenvalues as \(A\).

We first construct eigenfunctions corresponding to the eigenvalues of \(A\).
For simplicity, assume that \(A\) is diagonalizable over \(\mathbb C\), and let \(\bm{v}_i\) satisfy
\begin{align}
A\bm{v}_i=\mu_i\bm{v}_i .
\end{align}
We introduce the linear functions
\begin{align}
y_i(\bm{x}) := \bm{v}_i^\top\Sigma^{-1}\bm{x},
\end{align}
as a polynomial $P$.
Then, since \(\nabla y_i=\Sigma^{-1}\bm{v}_i\),
\begin{align}
(\tilde{A}\bm{x})\cdot \nabla y_i
=
\bm{x}^\top \tilde{A}^\top \Sigma^{-1}\bm{v}_i
=
\bm{x}^\top \Sigma^{-1}A\bm{v}_i
=
\mu_i y_i .
\end{align}
Because the second-order term in \eqref{eq_s:FP_conjugated_short} vanishes on linear functions, we obtain
\begin{align}
\mathcal L_{\mathrm{FP}}(y_i\rho_{\mathrm{ss}})
=
\mu_i\,y_i\rho_{\mathrm{ss}} .
\end{align}
Thus \(y_i(\bm{x})\rho_{\mathrm{ss}}(\bm{x})\) are eigenfunctions with eigenvalues \(\mu_i\).

Higher-order eigenvalues are obtained by examining the highest-degree part of a polynomial \(P\).
Let \(P_k\) be the homogeneous part of degree \(k\) of \(P\).
Since the linear functions \(y_i\) form a basis of linear polynomials, \(P_k\) is a linear combination of monomials
\begin{align}
y_{i_1}y_{i_2}\cdots y_{i_k},
\end{align}
where $i_j$ ($j \in{1, \dots, k}$) is the integer corresponding to the eigenvector $\boldsymbol{v}_i$. 
For such a monomial, the first-order term in \eqref{eq_s:FP_conjugated_short} gives
\begin{align}
(\tilde{A}\bm{x})\cdot \nabla
\left(
y_{i_1}\cdots y_{i_k}
\right)
=
(\mu_{i_1}+\cdots+\mu_{i_k})
\,y_{i_1}\cdots y_{i_k},
\end{align}
whereas the second-order term lowers the degree by two.
Therefore
\begin{align}
\mathcal L_{\mathrm{FP}}
\left(
y_{i_1}\cdots y_{i_k}\rho_{\mathrm{ss}}
\right)
=
(\mu_{i_1}+\cdots+\mu_{i_k})
\,y_{i_1}\cdots y_{i_k}\rho_{\mathrm{ss}}
+
\rho_{\mathrm{ss}}R(\bm{x}),
\end{align}
where \(R(\bm{x})\) is a polynomial of degree at most \(k-2\).

Finally, the lower-degree terms can be removed recursively.
Specifically, one can choose
\begin{align}
\widetilde P
=
y_{i_1}\cdots y_{i_k}
+
P_{k-2}
+
P_{k-4}
+\cdots ,
\end{align}
with each \(P_m\) homogeneous of degree \(m\), so that the lower-degree remainder cancels.
Because the degree decreases by two at each step, this procedure terminates after finitely many steps.
Hence $\widetilde P(\bm{x})\rho_{\mathrm{ss}}(\bm{x})$
satisfies
\begin{align}
\mathcal L_{\mathrm{FP}}(\widetilde P\rho_{\mathrm{ss}})
=
(\mu_{i_1}+\cdots+\mu_{i_k})\,\widetilde P\rho_{\mathrm{ss}}.
\end{align}

Therefore the eigenvalues of the Fokker--Planck operator are given by finite sums of the eigenvalues of \(A\),
\begin{align}
\lambda=\mu_{i_1}+\mu_{i_2}+\cdots+\mu_{i_k}.
\end{align}
In particular, the oscillatory frequencies associated with the Fokker--Planck spectrum are determined by the imaginary parts of the eigenvalues of the deterministic drift matrix \(A\).

}

{\color{black}
\section{Application of our decomposition to non-steady state dynamics}

\begin{figure}
    \centering
    \includegraphics[width=\linewidth]{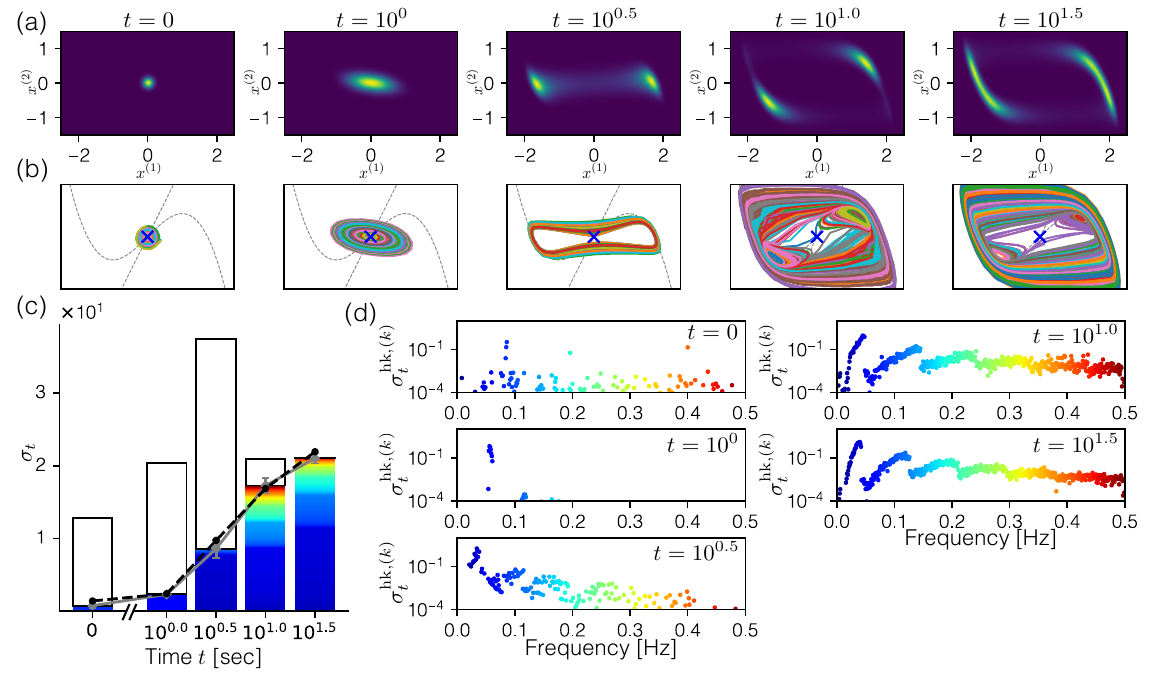}
    \caption{\color{black}
    Our decomposition enables us to analyze the time-dependent entropy production rate during a relaxation process from an initial Gaussian distribution centered at the origin to the stationary distribution.
    (a) Time evolution of the probability density $p_t(\bm{x})$ at representative times $t=0,\,10^0,\,10^{0.5},\,10^{1.0},\,10^{1.5}$.
    (b) Examples of trajectories of the virtual dynamics in Eq. \ref{eq_s:Langevin_hk_s} constructed from the instantaneous housekeeping local mean velocity field $\bm{\nu}_t^{\mathrm{hk}}$ at each time.
    The dashed curves indicate the nullclines of the deterministic drift.
    The blue cross denotes the fixed point of the deterministic drift.
    (c) Time-dependent behavior of the entropy production rate and its decomposition.
    The stacked bar plot shows the contributions from oscillatory modes $\sigma_t^{\mathrm{hk},(k)}$, where the colors represent frequencies $\chi_k$.
    The gray line indicates the sum of the contributions from the oscillatory modes, with error bars representing 95\% confidence intervals. 
    The black dashed line shows the true housekeeping entropy production rate. 
    The white portion represents the excess entropy production rate $\sigma_t^{\mathrm{ex}}$.
    (d) The contribution of each oscillatory mode to the housekeeping entropy production rate $\sigma_t^{\mathrm{hk},(k)}$ at different times $t$.
    Each panel corresponds to a representative time shown in (a).
    }
    \label{fig:nonst}
\end{figure}

We demonstrate that our decomposition can be applied to non-steady-state dynamics.
Specifically, we consider a relaxation process from an initial Gaussian distribution centered at the origin to the stationary distribution of the system.
Figure~\ref{fig:nonst}a shows the time evolution of the probability density $p_t(\bm{x})$ governed by the Fokker--Planck equation (Eq.~\ref{eq:Fokker--Planck}) at representative times.
When the initial distribution is chosen as a Gaussian centered at the origin, the density progressively deforms and approaches the stationary distribution as time increases.
Details of the numerical procedures and parameter settings used in this section are provided in Section \textit{``\nameref{sec:Methods_for_SI}''}.

To understand how the housekeeping entropy production rate $\sigma_t^\mathrm{hk}$ is generated during the relaxation process, we construct at each time the virtual dynamics driven by the instantaneous housekeeping local mean velocity field $\bm{\nu}_t^{\mathrm{hk}}$ as in Eq.~\ref{eq:Langevin_hk}.
Examples of trajectories generated by these virtual dynamics are shown in Fig.~\ref{fig:nonst}b.
Although the probability distribution $p_t(\bm{x})$ evolves in time and the system is therefore nonstationary, extracting the housekeeping component allows us to isolate the instantaneous sustaining directions of motion associated with the housekeeping entropy production rate.
In this way, even in a time-dependent setting, the virtual dynamics reveal the oscillatory structure responsible for the housekeeping contribution at each moment.
Notably, around $t \sim 10^{0.5}$, the probability distribution undergoes a pronounced change, and the qualitative structure of the virtual dynamics changes accordingly.

Figure~\ref{fig:nonst}c shows the time-dependent behavior of the total entropy production rate and its decomposition.
The stacked bar plot represents the contributions from oscillatory modes $\sigma_t^{\mathrm{hk},(k)}$, while the white portion corresponds to the excess entropy production rate $\sigma_t^{\mathrm{ex}}$.
The gray line indicates the sum of the oscillatory contributions, and the black dashed line shows the housekeeping entropy production rate.
Consistent with the structural change observed around $t=10^{0.5}$, the excess entropy production rate increases during this transient stage. 
As the system approaches the stationary distribution, the excess contribution gradually diminishes and eventually vanishes, leaving only the housekeeping component for sufficiently large $t$, where the distribution is close to stationarity.

The frequency-resolved contributions to the housekeeping entropy production rate $\sigma_t^{\mathrm{hk},(k)}$ at representative times are shown in Fig.~\ref{fig:nonst}d.
At early times, the contribution is dominated by a relatively limited set of frequencies.
Around $t \sim 10^{0.5}$, the pattern of contributions changes qualitatively.
For larger $t$, multiple oscillatory modes over a broader range of frequencies contribute to the housekeeping entropy production rate.
As the distribution further approaches stationarity, the frequency structure gradually stabilizes toward that characteristic of the steady-state dynamics.

These results demonstrate that our decomposition consistently captures the evolving oscillatory structure of entropy production even in nonstationary settings.
}

{\color{black}
\section{Finite sampling effects in low-noise metastable systems}

This section discusses how finite sampling affects the evaluation of our decomposition in low-noise metastable systems.
The main point is that, in the numerical setting considered in the present study, this issue does not constitute a fundamental limitation because the true underlying Langevin dynamics are assumed to be known.
Once the dynamics are specified, the stationary distribution and the local mean velocity field can be obtained directly from the discretized transition rate matrix on the state space grid, and additional trajectories can then be generated as needed.
Therefore, insufficient sampling can in principle be remedied by increasing the number of sampled trajectories and checking convergence with respect to sample size.

At the same time, low-noise metastable systems still require particular care when the decomposition is evaluated from a finite number $N$ of sampled trajectories.
If $N$ is too small, the sampled trajectories may not represent the full stationary dynamics sufficiently well, and the reconstructed entropy production can then be underestimated.
To illustrate this point, we consider a low-noise bistable FitzHugh--Nagumo system with parameters
$a=0.1, b=2.5, \tau=12.5, R=1, T=10^{-1}, I=0.$

Figure~\ref{fig:bistable}a shows example trajectories of the virtual dynamics for this parameter regime.
The trajectories consist only of small motions around the two stable fixed points.
As a result, when the number of sampled trajectories is too small, the sample can easily become biased toward one basin.
This sampling bias affects the evaluation of the decomposition, even though the underlying stationary distribution and local mean velocity field themselves are known exactly in the present numerical framework.

\begin{figure}[H]
    \centering
    \includegraphics[width=0.5\linewidth]{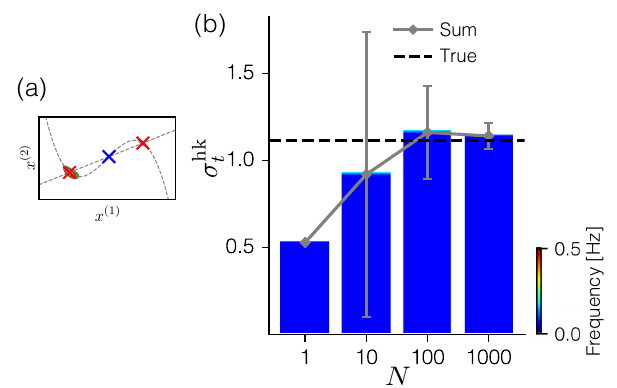}
    \caption{
    \color{black}
    (a) Examples of trajectories of the virtual dynamics in Eq.~6 for the noisy FitzHugh--Nagumo model in Eq.~13. 
    The black dashed line represents the nullclines of the noisy FitzHugh--Nagumo model, which were calculated from the original Langevin dynamics in Eq.~13 by ignoring the noise term. 
    The blue and red crosses represent the unstable and stable fixed points, respectively.
    (b) The $N$-dependent behavior of the housekeeping entropy production rate and its decomposition, where $N$ represents the number of trajectories used to approximate the expectation in the calculation of our decomposition. 
    The stacked bar plot shows the sum of the contributions from the oscillatory modes. 
    The colors represent the frequencies of the oscillatory modes. 
    The gray line indicates the sum of the contributions from the oscillatory modes, with error bars representing 95\% confidence intervals. 
    The black dashed line shows the true housekeeping entropy production rate. 
    }
    \label{fig:bistable}
\end{figure}

Figure~\ref{fig:bistable}b shows how the housekeeping entropy production reconstructed from the decomposition depends on the number of sampled trajectories $N$.
When $N$ is small, the sum of the decomposed contributions underestimates the true housekeeping entropy production rate.
As $N$ is increased, the reconstructed value approaches the true value.
At the same time, the confidence intervals become progressively narrower, indicating improved statistical reliability.
For $N=1$, the confidence interval is not shown because it is not defined.

These results show that finite sampling in low-noise metastable systems can cause underestimation of the reconstructed entropy production when too few trajectories are used.
However, in the present numerical setting this is not a fundamental obstacle, because the true dynamics are known and additional trajectories can be generated as needed.
Accordingly, the discrepancy can be reduced in practice by taking $N$ sufficiently large and checking convergence with respect to sample size.

}

{\color{black}
\section{Supplementary Methods}
\phantomsection
\makeatletter
\def\@currentlabelname{Supplementary Methods}
\makeatother
\label{sec:Methods_for_SI}

This section describes the numerical procedures used for the noisy FitzHugh–Nagumo model (Sec.~\textit{``\nameref{sec:fitzhugh}''}).
The numerical calculations consist of the following steps:
(i) computing the housekeeping part of the local mean velocity $\bm{\nu}_t^\mathrm{hk}$ in the steady state,
(ii) simulating the virtual dynamics driven by $\bm{\nu}_t^\mathrm{hk}$ in Eq.~\ref{eq:Langevin_hk}, and
(iii) extracting Koopman eigenfunctions and modes from the generated time-series data and calculating the terms of our decomposition in Eq.~\ref{eq:decomposition_EP}.
The numerical experiments were conducted under the following parameter settings for the noisy FitzHugh–Nagumo model.
For Fig.~\ref{fig:FN_demo}a–e, we used $a=0$, $b=0.5$, $I=0$, $T=10^{-3}$, and $\tau=12.5$.
For Fig.~\ref{fig:FN_demo}f–h, the same parameters were used except that $\tau$ was varied from $2.5$ to $22.5$ in increments of $5$.
For Fig.~\ref{fig:FN_I}, we set $a=0$, $b=0.5$, $T=10^{-3}$, and $\tau=12.5$, while varying $I$ from $0$ to $2.5$ in increments of $0.1$.
For Fig.~\ref{fig:FN_T}, we fixed $a=0$, $b=2$, $I=0$, and $\tau=12.5$, with $T$ varied from $10^{-4}$ to $10^{-1}$ in logarithmic steps of $0.1$.

\subsection{Calculation of the local mean velocity}
We calculated the housekeeping part of the local mean velocity $\bm{\nu}_t^\mathrm{hk}$ from the noisy FitzHugh--Nagumo model in Eq.~\ref{eq:Langevin_FN}. 
The decomposition was performed in the steady state, where the excess part vanishes, and the local mean velocity $\bm{\nu}_t$ coincides with its housekeeping part $\bm{\nu}_t^\mathrm{hk}$.
Thus, determining the local mean velocity $\bm{\nu}_t$ is sufficient to obtain its housekeeping part $\bm{\nu}_t^\mathrm{hk}$.
To compute the local mean velocity $\bm{\nu}_t$ at the steady state, we estimated the steady-state distribution $p_t$, which satisfies $\pdv{t} p_t(\bm{x})=0$, using a discretization approach following the method in Ref. \citeSI{gingrich2017inferring_}: the continuous Langevin equation in Eq.~\ref{eq:Langevin_FN} was converted into a discrete transition rate matrix by dividing the state space into a grid.
To discretize the state space, we divided the two variables of the FitzHugh–Nagumo model, $x^{(1)}$ and $x^{(2)}$, into a grid. The variable $x^{(1)}$ was discretized into $10^4$ intervals ranging from $-5$ to $5$, and $x^{(2)}$ was discretized into $10^4$ intervals spanning $-5+I$ to $5+I$.
This resulted in a total of $10^{4\times 2}$ grid points.
In this discrete system with $10^{4\times 2}$ states, we constructed the transition rate matrix corresponding to the Langevin equation in Eq.~\ref{eq:Langevin_FN}.
The steady-state distribution for the grid points was then obtained by computing the eigenvector corresponding to the zero eigenvalue of the transition rate matrix.
From this steady-state distribution, we computed the gradient $\nabla \ln p_t$ by interpolating $p_t$ between grid points using cubic-spline interpolation and differentiating the resulting cubic polynomial.
This method allowed us to compute the local mean velocity not only at the grid points but also at arbitrary locations within the continuous state space.

\subsection{Simulating the virtual dynamics}
Having obtained the housekeeping part of the local mean velocity $\bm{\nu}_t^\mathrm{hk}$, we generated time-series trajectories by simulating the virtual dynamics driven by $\bm{\nu}_t^\mathrm{hk}$ in Eq.~(\ref{eq:Langevin_hk}). 
We simulated the dynamics using an eighth-order Runge–Kutta method with a time step $\Delta s = 1$, generating trajectories of length $S=150$ steps. 
For each parameter setting of the noisy FitzHugh–Nagumo model, we generated $N=1000$ independent trajectories to evaluate the decomposition of the entropy production rate. 
These trajectories serve as Monte Carlo samples for approximating the terms in our decomposition. 
The initial conditions for the trajectories were sampled from the discretized steady-state distribution $p_t(\bm{x})$. 
Hereafter, we denote by $\bm{x}_{n,s}$ the state at time $s$ of the $n$-th trajectory.

\subsection{Extraction of Koopman eigenfunctions and modes} 
From the simulated time-series data, we estimated the Koopman eigenfunctions $\{\phi_{n, k}\}_{k=1}^{r_n}$, eigenvalues $\{\lambda_{n, k}\}_{k=1}^{r_n}$, and modes $\{\bm{v}_{n, k}\}_{k=1}^{r_n}$ for each trajectory.  
Because different trajectories have different supports in state space, the eigenfunctions obtained from distinct trajectories were regarded as different functions.  
Here, $r_n$ denotes the number of extracted modes for the $n$-th trajectory, and the double subscript $_{n,k}$ indicates the $k$-th eigenfunction, eigenvalue, or mode associated with that trajectory. 
{\color{black}The number of extracted modes $r_n$ varies across trajectories, and the procedure for determining $r_n$ is described later.}
To obtain these quantities, we employed Hankel DMD~\citeSI{tu2014on_, arbabi2017ergodic_, brunton2017chaos_} in combination with physics-informed DMD (PiDMD)~\citeSI{baddoo2023physics_} via the PyDMD Python package~\citeSI{ichinaga2024pydmd_}.  
Hankel DMD constructs a vector of $h_n$ observable functions, $\bm{g}(\bm{x})=(\bm{g}_1(\bm{x})^\top, \bm{g}_2(\bm{x})^\top, \dots, \bm{g}_{h_n}(\bm{x})^\top)^\top=(\mathrm{Id}(\bm{x})^\top, (e^{\Delta s\mathcal{K}}\mathrm{Id}(\bm{x}))^\top, \dots, (e^{({h_n}-1)\Delta s\mathcal{K}}\mathrm{Id}(\bm{x}))^\top)^\top\in \mathbb{R}^{dh_n}$, where $\mathrm{Id}(\bm{x})=\bm{x}$ is the identity function.  
Here, $h_n$ denotes the number of time delays chosen for the $n$-th trajectory, which serves as a hyperparameter of the fitting procedure.  
{\color{black}The procedure for selecting $h_n$ is also described later.}
This vector is obtained by a time-delay embedding of the time-series data, i.e., $\bm{g}(\bm{x}_{n,s})=(\bm{x}_{n,s}^\top, \bm{x}_{n, s+\Delta s}^\top, \dots, \bm{x}_{n, s+({h_n}-1)\Delta s}^\top)^\top$.  
Since the Koopman generator $\mathcal{K}$ is linear, the time evolution of this observable vector can be approximated by a linear dynamical system, even when the dynamics of $\bm{x}_{n,s}$ are nonlinear.

To estimate the Koopman generator $\mathcal{K}$ while ensuring that its eigenvalues are purely imaginary, we applied PiDMD, which constrains the representation matrix of $e^{\Delta s\mathcal{K}}$ to be unitary during the fitting procedure. 
Let $L$ denote the representation matrix of $e^{\Delta s\mathcal{K}}$. 
Under stationarity, the covariance matrix $\Sigma$ of the observable vectors satisfies $\Sigma = L\Sigma L^*$, which implies that when $\Sigma$ is the identity matrix, $L$ must be unitary.
Therefore, before applying PiDMD, we linearly transformed the delay-embedded data so that its covariance matrix became the identity matrix, ensuring that the fitted $L$ satisfies the unitarity condition under stationarity.
Specifically, we first centered the delay-embedded observable vectors by subtracting their temporal mean, $\overline{\bm{g}} = (1/(S-h_n+1))\sum_{s=1}^{S-h_n+1} \bm{g}(\bm{x}_{n,s})$, $G = \qty(\bm{g}(\bm{x}_{n,1})-\overline{\bm{g}}, \ \bm{g}(\bm{x}_{n,2})-\overline{\bm{g}},\ \dots, \ \bm{g}(\bm{x}_{n,S-h_n+1})-\overline{\bm{g}})$ 
and then performed a singular value decomposition $G = U S V^\top$. 
We whitened the data as $\Tilde{G} = S^{-1} U^\top G$, and applied PiDMD to the new data matrix $\Tilde{G}$ to obtain the Koopman modes $\{\Tilde{\bm{v}}_{n,k}\}$. 
{\color{black}Here, the number of extracted modes $r_n$ was determined for each trajectory from this singular value decomposition by applying optimal singular value hard thresholding (SVHT)~\citeSI{gavish2014optimal_} to the singular values of $G$.}
Finally, the obtained modes were mapped back to the original coordinate system as $\bm{v}_{n,k} = U S \Tilde{\bm{v}}_{n,k}$, and the temporal mean $\overline{\bm{g}}$ was added back to the reconstructed trajectories so that they could be interpreted in the original observable space $\bm{g}(\bm{x}_{n,s})$.

For each trajectory, the number of time delays $h_n$ was chosen from the range of 1 to 100 in increments of 1 to minimize the reconstruction error based on Eq.~\ref{eq_s:oscillation_x}, $\sum_{s=1}^S \left\|\bm{x}_{n,s} - \sum_{k=1}^{r_n} e^{\lambda_{n,k} s}\,\phi_{n,k}(\bm{x}_{n,0})\,\bm{v}_{n,k} \right\|^2$, which varies with the delay dimension $h_n$ because the extracted eigenvalues $\lambda_{n,k}$, eigenfunctions $\phi_{n,k}$, and modes $\bm{v}_{n,k}$ depend on the chosen $h_n$ during the fitting procedure.

\subsection{Computation of the terms of our decomposition}
From the Koopman eigenvalues, eigenfunctions, and modes obtained for each trajectory, we computed the terms of our decomposition as follows:
\begin{align}
    &\chi_{n,k} = |\lambda_{n,k}/(2\pi \mathrm{i})|\\
    &J_{n,k} = \frac{1}{S}\sum_s \qty(\phi_{n, k}\qty(\bm{x}_{n, s})\bm{v}_{n, k})^* D_t^{-1}\qty(\phi_{n, k}\qty(\bm{x}_{n, s})\bm{v}_{n, k}), \\
    &\sigma_t^{\mathrm{hk}, (n, k)} = (2\pi)^2\chi_{n, k}^2J_{n,k},
\end{align}
where the expected values in Eq.~\ref{eq_s:decomposition_EP} are approximated as the time-average within the trajectory.
Since each eigenfunction is supported only within its corresponding trajectory and does not overlap with those from other trajectories, 
the associated quantities $\chi_{n,k}, J_{n,k}$, and $\sigma_t^{\mathrm{hk}, (n,k)}$ were considered as distinct terms for each trajectory. 
Let $r=\sum_{n=1}^N r_n$ denote the total number of extracted modes across all trajectories. 
To summarize the results in a form consistent with ensemble expectations, we concatenated these quantities and reindexed them, dividing each trajectory-specific contribution by $N$:
$\{J_k\}_{k=1}^r = \{\, J_{1,1}/N, ..., J_{1,r_1}/N, ..., J_{N,1}/N, ..., J_{N,r_N}/N \,\}$, 
$\{\sigma_t^{\mathrm{hk}, (k)}\}_{k=1}^r=\{\sigma_t^{\mathrm{hk}, (1, 1)}/N, ..., \sigma_t^{\mathrm{hk}, (1, r_1)}/N, ..., \sigma_t^{\mathrm{hk}, (N, 1)}/N, ..., \sigma_t^{\mathrm{hk}, (N, r_N)}/N\}$.
These reindexed quantities were then plotted in Figs.~\ref{fig:FN_demo},~\ref{fig:FN_I}, and~\ref{fig:FN_T}.  
In the frequency-resolved panels in Figs.~\ref{fig:FN_demo}d, g, h;~\ref{fig:FN_I}c;~\ref{fig:FN_T}d, we applied a moving-window procedure to enable clearer comparison across frequencies: this both alleviates overlap among points and compensates for non-uniform point densities across frequency bands. 
Within each bin of width $10^{-3}$, the values were summed and plotted at the bin center.

\subsection{Estimation of confidence intervals}
To assess the finite-sample variability of the sum of the decomposition, we treated $\{\sum_{k=1}^{r_n} \sigma_t^{\mathrm{hk}, (n, k)}\}_{n=1}^N$ as $N$ independent samples.  
For clarity, we define the trajectory-wise sum of the decomposition as 
$\sigma_n^\mathrm{sum} = \sum_{k=1}^{r_n} \sigma_t^{\mathrm{hk}, (n, k)}$.  
The sample mean 
$\bar\sigma^\mathrm{sum} = \tfrac{1}{N}\sum_{n=1}^N \sigma_n^\mathrm{sum}$ 
and the standard error 
$\mathrm{SE}^\mathrm{sum} = \sqrt{\tfrac{1}{N(N-1)}\sum_{n=1}^N (\sigma_n^\mathrm{sum} - \bar\sigma^\mathrm{sum})^2}$ 
were then computed.  
From these, 95\% confidence intervals were constructed as
\begin{align}
    \bar\sigma^\mathrm{sum} \pm t_{N-1,\,0.975}\,\mathrm{SE}^\mathrm{sum},
\end{align}
where $t_{N-1,0.975}$ denotes the $97.5\%$ quantile of the $t$-distribution with $N-1$ degrees of freedom.  
These confidence intervals are shown as error bars for the sum of the decomposition in Figs.~\ref{fig:FN_demo}–\ref{fig:FN_T}.

\subsection{Calculation of the true values of the housekeeping entropy production rates}
To validate that the sum of our decomposition recovers the true value, we computed the true housekeeping entropy production rate using the method of Ref.~\citeSI{gingrich2017inferring_}. 
Specifically, with the steady-state distribution $p_t(\bm{x})$ and the local mean velocity field $\bm{\nu}_t^\mathrm{hk}(\bm{x})$ obtained as described above, the true housekeeping entropy production rate was calculated, following the definition in Eq.~\ref{eq:EP_hk}, by numerically integrating $\bm{\nu}_t^\mathrm{hk}(\bm{x})^\top D_t^{-1}\bm{\nu}_t^\mathrm{hk}(\bm{x}) \, p_t(\bm{x})$. 
This result is shown in Fig.~\ref{fig:FN_demo}e and as the black dashed lines in Figs.~\ref{fig:FN_demo}f, \ref{fig:FN_I}b, and \ref{fig:FN_T}c.

\subsection{Calculation of correlation times}
The correlation time $\tau_{\mathrm{corr}}$ in Fig. \ref{fig:FN_T}b represents the degree of temporal coherence in dynamical systems, and serves as a signature of {\color{black} coherent resonance}~\citeSI{pikovsky1997coherence_}. It is theoretically defined as
\begin{align}
    \tau_{\mathrm{corr}} &= \int_0^\infty C(u)^2 \, du, \label{eq_s:corr_time}
\end{align}
where $C(u):=\text{Cov}[x^{(1)}_{t'}, x^{(1)}_{t'+u}]/\text{Var}[x_{t'}^{(1)}]$ represents the autocorrelation function of $x_t^{(1)}$ with the time lag of $u$. 
Here, $\text{Cov}[x^{(1)}_{t'}, x^{(1)}_{t'+u}]$
is the covariance between $x^{(1)}_{t'}$ and $x^{(1)}_{t'+u}$ in the steady state, and $\text{Var}[x^{(1)}]$ is the variance of $x^{(1)}$ in the steady state.
Because the process is in the steady state, these quantities do not depend on the choice of the reference time $t'$, but only on the lag $u$. 
Since $\tau_{\mathrm{corr}}$ integrates the squared autocorrelation over time, a higher value indicates that correlations decay more slowly, corresponding to more temporally ordered dynamics.

The correlation times were numerically calculated as follows.  
We simulated $N_\text{sr}=1000$ independent trajectories $\{\bm{x}_{n,t}\}_{t=1}^{S_\text{sr}}$ of the Langevin dynamics in Eq.~\ref{eq:Langevin_FN} by the Euler–Maruyama method, with a step size $\Delta t = 10^{-2}$ and $S_\text{sr}=10^4$ steps for each trajectory $n$.  
The initial conditions of these simulations were sampled from the stationary distribution of Eq.~\ref{eq:Langevin_FN}, which was also used in the numerical calculation of the decomposition.  
Note that these trajectories differ from those following the virtual deterministic process in Eq.~\ref{eq:Langevin_hk} used in the decomposition analysis.  
From the first component $x_{n,t}^{(1)}$ of each trajectory, we computed the autocorrelation function at lag $\ell \Delta t$ $(0 \le \ell \le S_\text{sr}-1,\ \ell \in \mathbb{Z}_{\ge0})$ as
\begin{align}
    C_n(\ell \Delta t) 
    &= \frac{\frac{1}{S_\text{sr}-\ell}\sum_{j=1}^{S_\text{sr}-\ell} 
        \bigl(x_{n, j\Delta t}^{(1)} - \bar{x}_n^{(1)} \bigr)
        \bigl(x_{n,(j+\ell)\Delta t}^{(1)} - \bar{x}_n^{(1)} \bigr)}
      {\frac{1}{S_\text{sr}}\sum_{j=1}^{S_\text{sr}} \left(x_{n,j\Delta t}^{(1)} - \bar{x}_n^{(1)}\right)^2},
\end{align}
where $\bar{x}_n^{(1)} = \tfrac{1}{S_\text{sr}} \sum_{j=1}^{S_\text{sr}} x_{n,j\Delta t}^{(1)}$ is the time average within trajectory $n$.  
This estimator of the autocorrelation is motivated by the fact that, in the steady state, the variance does not depend on the time index, and the covariance depends only on the lag $\ell \Delta t$ but not on the reference time $j\Delta t$.  
The trajectory-specific correlation time was then obtained by summing the squared autocorrelation over lags,
\begin{align}
    \tau_{\mathrm{corr}, n} = \sum_{\ell=0}^{S_\text{sr}-1} C_n(\ell \Delta t)^2 \,\Delta t.
\end{align}
The mean correlation time across trajectories was calculated as $\tau_\mathrm{corr} = N_\text{sr}^{-1}\sum_{n=1}^{N_\text{sr}} \tau_{\mathrm{corr}, n}$ and plotted in Fig.~\ref{fig:FN_T}b, with error bars indicating the 95\% confidence interval.  
The confidence intervals were constructed by evaluating the variability among $\{\tau_{\mathrm{corr}, n}\}_{n=1}^{N_\text{sr}}$ using the $t$-distribution with $N_\text{sr}-1$ degrees of freedom.

\subsection{Comparison with the limit-cycle frequency}
For the results shown in Fig.~\ref{fig:linear_stability_analsis}b-d, we estimated the oscillation frequency directly from simulations of the noisy FitzHugh--Nagumo model in Eq.~\ref{eq:Langevin_FN}. 
The parameter values were set to
$a = 0$, $b = 0.5$, $I = 0$, $T = 10^{-4}$, and $\tau = 12.5$.
Under this parameter setting, the deterministic part of the dynamics exhibits limit-cycle oscillations, and the stochastic trajectories fluctuate around this cycle.

To estimate the oscillation frequency, we simulated $N_{\mathrm{lc}}=1000$ independent trajectories of the noisy FitzHugh--Nagumo dynamics using the Euler--Maruyama method with time step $\Delta t = 10^{-2}$ for $S_{\mathrm{lc}}=10^5$ steps. 
We denote the state of the $n$-th trajectory at step $s$ by
\(
\bm{x}_{n,s}=(x^{(1)}_{n,s},x^{(2)}_{n,s})
\).

For each trajectory, we introduced a phase variable defined from the two-dimensional state,
\begin{align}
\theta_{n,s}
=
\arg\!\left(
(x^{(1)}_{n,s}-\bar{x}^{(1)}_n)
+
i(x^{(2)}_{n,s}-\bar{x}^{(2)}_n)
\right),
\end{align}
where
\(
\bar{x}^{(j)}_n = \tfrac{1}{S_{\mathrm{lc}}}\sum_{s=1}^{S_{\mathrm{lc}}} x^{(j)}_{n,s}
\)
denotes the temporal mean of the $j$-th component along trajectory $n$.
The phase sequence $\{\theta_{n,s}\}$ was unwrapped to remove discontinuities at multiples of $2\pi$.

The oscillation frequency for trajectory $n$ was then estimated from the long-time average rate of phase increase.
Specifically, we computed the average amount by which the phase advances per unit time along the trajectory,
\begin{align}
\omega_n
=
\frac{\theta_{n,S_{\mathrm{lc}}}-\theta_{n,1}}
{(S_{\mathrm{lc}}-1)\Delta t},
\end{align}
which represents the mean angular velocity of the trajectory.
The corresponding oscillation frequency was then obtained as
\begin{align}
f_n = \frac{\omega_n}{2\pi}.
\end{align}

Finally, the limit-cycle frequency used for comparison with the dominant oscillatory mode of the housekeeping entropy production rate was obtained by averaging over trajectories,
\begin{align}
f_{\mathrm{LC}}
=
\frac{1}{N_{\mathrm{lc}}}
\sum_{n=1}^{N_{\mathrm{lc}}} f_n .
\end{align}

\subsection{Comparison with the frequencies extracted by linear stability analysis}

For the comparison with the frequencies predicted by linear stability analysis in Fig.~\ref{fig:oscillation_main} and Fig.~\ref{fig:linear_stability_analsis}, we computed the frequency determined by the fixed-point analysis of the virtual dynamics driven by the housekeeping local mean velocity field $\bm{\nu}_t^\mathrm{hk}$.

Using the velocity field $\bm{\nu}_t^\mathrm{hk}$ obtained from the discretized state space and its interpolation as described above, the fixed point $\bm{x}_{\mathrm{fix}}$ of the virtual dynamics was obtained by minimizing $\|\bm{\nu}_t^\mathrm{hk}(\bm{x})\|^2$, which identifies the point where $\bm{\nu}_t^\mathrm{hk}(\bm{x}_{\mathrm{fix}})=\bm{0}$. 
After locating the fixed point, the Jacobian $ \partial\bm{\nu}_t^\mathrm{hk}/ \partial\bm{x} \big|_{\bm{x}=\bm{x}_{\mathrm{fix}}}$ was evaluated numerically using a central finite-difference approximation with step size $h=10^{-6}$.

The frequency predicted by linear stability analysis was then obtained from 
\begin{align}
f_{\mathrm{LSA}}
=
\frac{1}{2\pi}\left|
\mathrm{Im}(\lambda_{\mathrm{LSA}}
)\right|,
\end{align}
where $\lambda_{\mathrm{LSA}}$ is an eigenvalue of the Jacobian $\partial \bm{\nu}_t^\mathrm{hk}/ \partial\bm{x} \big|_{\bm{x}=\bm{x}_{\mathrm{fix}}}$ .
In the present two-dimensional system, if the eigenvalues have nonzero imaginary parts, they form a complex-conjugate pair, so either eigenvalue gives the same absolute value of the imaginary part.
This frequency was used for comparison with the dominant oscillatory mode extracted by our decomposition in Fig.~\ref{fig:oscillation_main} and Fig.~\ref{fig:linear_stability_analsis}.

\subsection{Application of our decomposition to non-steady state dynamics}
For the results shown in Fig.~\ref{fig:nonst}, we analyzed a non-steady-state relaxation process of the noisy FitzHugh--Nagumo model in Eq.~\ref{eq:Langevin_FN} with parameters $a=0$, $b=0.5$, $I=0$, $T=10^{-3}$, and $\tau=12.5$. 
The initial distribution was chosen as a Gaussian distribution
$
p_0(\bm{x}) =
\mathcal{N}\!\left(
\mqty(0 \\ 0),
\mqty(0.1 & 0 \\ 0 & 0.1)
\right).
$
The time evolution of the probability distribution was first computed on a discretized state space. 
At representative time points during the relaxation process (Fig.~\ref{fig:nonst}a), we evaluated the instantaneous velocity field and decomposed it into excess and housekeeping components. 
Using the resulting housekeeping velocity field $\bm{\nu}_t^{\mathrm{hk}}$, we constructed the corresponding virtual dynamics and generated $N=1000$ trajectories of length $S=150$ steps from initial conditions sampled from the instantaneous distribution $p_t(\bm{x})$. 
The Koopman-mode decomposition was then applied to these trajectories to compute the frequency-resolved contributions to the housekeeping entropy production rate in Fig.~\ref{fig:nonst}c-d. 

To compute the time evolution of the probability distribution, the state space was discretized on a uniform grid over $[-5,5]\times[-5,5]$ with $2000\times2000$ grid points, and the corresponding transition-rate matrix $W$ was constructed in the same manner as in the steady-state analysis. 
Let $\bm{P}_t$ denote the vector obtained by flattening the discretized probability distribution on the grid. 
Starting from the initial vector $\bm{P}_0$, the probability distribution was evolved according to
\begin{align}
\pdv{\bm{P}_t}{t} = W\bm{P}_t,
\end{align}
which corresponds to the discretized Fokker–Planck equation. 
This equation was solved using the stiff ODE solver \texttt{ode15s}, and the probability distribution was evaluated at the representative times shown in Fig.~\ref{fig:nonst}a.

For each time point, the instantaneous velocity field was decomposed into excess and housekeeping components. 
To obtain the excess component, we introduced a scalar potential $\phi(\bm{x})$ and defined the excess velocity field as the gradient flow $\bm{\nu}_t^{\mathrm{ex}}=\nabla \phi$. 
The potential was determined so that the continuity equation
\begin{align}
\frac{\partial p_t(\bm{x})}{\partial t}
=
-
\nabla \cdot
\left(
p_t(\bm{x}) \nabla \phi(\bm{x})
\right)
\end{align}
is satisfied, where $\bm{x}=(x^{(1)},x^{(2)})$. 
On the discretized grid, let $\bm{\phi}$ denote the vector obtained by evaluating $\phi(\bm{x})$ at the grid points. 
Introducing finite-difference operators $G_1$ and $G_2$ that approximate the spatial derivatives $\partial/\partial x^{(1)}$ and $\partial/\partial x^{(2)}$, the discretized form of the above continuity equation can be written as
\begin{align}
\pdv{\bm{P}_t}{t}
=
-
\left(
G_1^\top \mathrm{diag}(\bm{P}_t) G_1
+
G_2^\top \mathrm{diag}(\bm{P}_t) G_2
\right)
\bm{\phi}.
\end{align}
Here $\mathrm{diag}(\bm{P}_t)$ denotes the diagonal matrix whose diagonal elements are given by the components of $\bm{P}_t$. 
This equation corresponds to the discretized form of the continuity equation above. 
We solved this linear equation for $\bm{\phi}$ using a least-squares procedure. 
The excess velocity field on the grid was then obtained from
\begin{align}
\nu_t^{\mathrm{ex},(1)} = G_1 \bm{\phi},
\qquad
\nu_t^{\mathrm{ex},(2)} = G_2 \bm{\phi}.
\end{align}
The excess entropy production rate shown in Fig.~\ref{fig:nonst}c was computed numerically on the grid using this excess velocity field together with the discretized probability distribution. 
The housekeeping component was then defined as the residual
\begin{align}
\bm{\nu}_t^{\mathrm{hk}}
=
\bm{\nu}_t
-
\bm{\nu}_t^{\mathrm{ex}} .
\end{align}

At $t=10^{0.5}$, where the probability distribution changes rapidly, the numerical accuracy of the velocity field becomes more sensitive to the grid resolution. 
For this time point only, after computing the distribution on the $2000\times2000$ grid, it was refined to a $3000\times3000$ grid by cubic-spline interpolation before constructing the virtual dynamics. 
After simulating $N=1000$ trajectories from the resulting velocity field, we excluded trajectories that did not return sufficiently close to their initial points during the final part of the orbit. 
Specifically, for each trajectory we computed the minimum squared distance between the initial point and the last 30 time steps, and retained only those satisfying 
\begin{align}
\min_{s\in\{S-29,\dots,S\}}
\|\bm{x}_{n,s}-\bm{x}_{n,0}\|^2 < 10^{-4}.
\end{align}

For each time point, the Koopman eigenvalues, eigenfunctions, and modes were extracted from the virtual trajectories in the same manner as in the steady-state analysis. 
From these quantities, we computed the frequency-resolved contributions $\sigma_t^{\mathrm{hk},(n,k)}$ and aggregated them across trajectories. 
The quantities plotted in Fig.~\ref{fig:nonst}d correspond to these frequency-resolved contributions. 
The stacked bars in Fig.~\ref{fig:nonst}c were obtained by summing the housekeeping contributions over oscillatory modes and adding the excess entropy production rate.

}



\providecommand{\noopsort}[1]{}\providecommand{\singleletter}[1]{#1}%

}


\end{document}